# Tracking Carbapenem-Resistant Pathogens in Hospital Wastewater: the focus on *Acinetobacter baumannii* and *Pseudomonas aeruginosa*

**Magdalena Męcik [1], Kornelia Stefaniak [1], Monika Harnisz [1], Ewa Felis [2,3], Sylwia Bajkacz [3,4], Joanna Wilk[4], Karolina Dudek[1] and Ewa Korzeniewska [1*]**

[1] Department of Water Protection Engineering and Environmental Microbiology, Faculty of Geoengineering, University of Warmia and Mazury in Olsztyn, Prawocheńskiego 1, 10-720 Olsztyn, Poland

[2] Silesian University of Technology, Faculty of Energy and Environmental Engineering, Environmental Biotechnology Department, Akademicka 2, Gliwice, 44-100, Poland

[3] Silesian University of Technology, Biotechnology Center, B. Krzywoustego 8, Gliwice, 44-100, Poland

[4] Silesian University of Technology, Faculty of Chemistry, Department of Inorganic Chemistry, Analytical Chemistry and Electrochemistry, B. Krzywoustego 6, Gliwice, 44-100, Poland

* Corresponding author: ewa.korzeniewska@uwm.edu.pl

**Abstract**

Carbapenem-resistant *Pseudomonas aeruginosa* (CRPA) and *Acinetobacter baumannii* (CRAB) represent a major clinical and epidemiological challenge and pose a growing threat to public health and the environment. Accordingly, CRPA and CRAB were investigated in hospital wastewater (HWW) collected during winter and summer 2024 from 64 healthcare facilities across all 16 Polish voivodeships. To our knowledge, this study constitutes the first nationwide, large-scale assessment in Poland of carbapenem resistance in these high-risk pathogens in hospital wastewater. The study aimed to determine the prevalence of carbapenem-resistant bacteria (CRB) in HWW discharged into the public sewer system and municipal wastewater treatment plants (WWTPs). In addition, associations between CRB prevalence, hospital geographic location, and sampling season were analyzed to identify spatial and temporal patterns of carbapenem resistance (CR). Carbapenem-resistant *P. aeruginosa* were predominant in all studied regions. Carbapenem-resistant *A. baumannii* were identified in a smaller percentage of samples and were characterized by greater genotypic diversity. The ERIC-PCR assay confirmed the presence of both closely related strains and unique genetic profiles, which suggests that CRB emissions into the environment have a complex character. The statistical analysis revealed significant relationships between CRB counts, the physicochemical parameters of HWW, and antibiotic concentrations in HWW samples. In addition, the tested samples harbored many antibiotic resistance genes (ARGs), which confirms that HWW is a significant reservoir of mobile genetic elements (MGEs) involved in the spread of antibiotic resistance. The results of the study indicate that HWW should be rigorously monitored and managed to minimize risks to public health and environment.

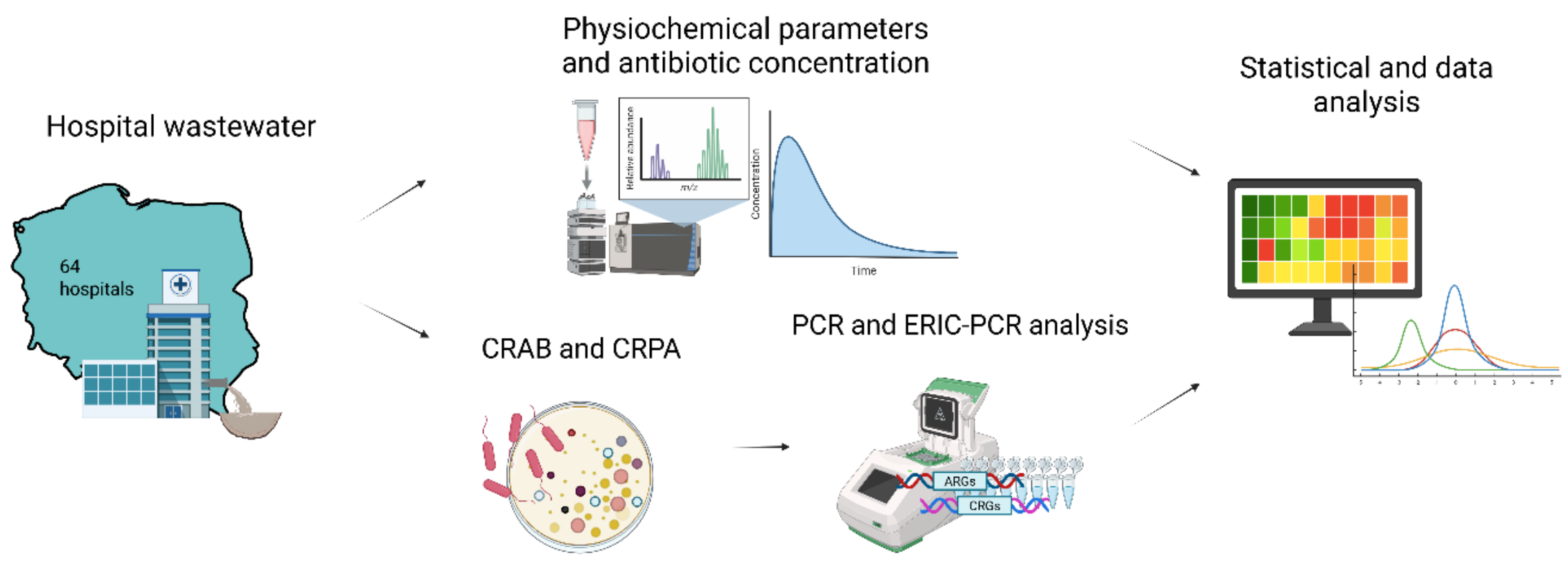

**Abbreviations**

**ACCAH** Annual Consumption of Carbapenem Antibiotics in Hospitals
**ACRC** All Carbapenem Resistant Colonies
**ARB** Antibiotic Resistant Bacteria
**ARGs** Antibiotic Resistance Genes
**CCRC** Cream Carbapenem Resistant Colonies
**CR** Carbapenem Resistant
**CRAB** Carbapenem Resistant *Acinetobacter baumannii*
**CRB** Carbapenem Resistant Bacteria
**MDR** Multidrug-resistant
**MGEs** Mobile Genetic Elements
**XDR** Extensively Drug-Resistant
**CRGs** Carbapenem Resistance Genes
**CRPA** Carbapenem Resistant *Pseudomonas aeruginosa*
**HWW** Hospital Wastewater
**VWW** Volume of Evacuated Wastewater
**WWTPs** Wastewater Treatment Plants

## 1. Introduction

Growing levels of bacterial resistance to antibiotics, including carbapenems, pose one of the greatest global health challenges. According to estimates, antibiotic resistance contributed to around 4.95 million global deaths in 2019 (World Health Organization 2024). The number of scientific reports on antimicrobial resistance, including carbapenem resistance (CR), antibiotic resistance genes (ARGs), and carbapenem resistance genes (CRGs) continues to increase each year (Jean et al., 2022). Special attention is given to pathogenic multidrug-resistant (MDR) and extensively drug-resistant (XDR) strains harboring genes that encode resistance to various classes of antibiotics, which pose a significant public health threat, both at present and in the foreseeable future (Catalano et al., 2022).

According to the World Health Organization (WHO), such bacteria include carbapenem-resistant *Acinetobacter baumannii* (CRAB), classified as critical pathogens, and carbapenem-resistant *Pseudomonas aeruginosa* (CRPA), classified as high priority pathogens (World Health Organization 2024). The spread of CRAB poses a significant global challenge in the context of public health and environmental protection. These highly virulent strains are resistant to antibiotics, which limits the availability of effective treatment options, contributes to severe nosocomial infections that are particularly dangerous for patients in intensive care units and have been linked to alarmingly high mortality rates in recent years (Jiang et al., 2022). Carbapenem-resistant *P. aeruginosa* are also frequently identified in hospitals and pose a particular threat to immunocompromised patients and patients in intensive care units (Centers for Disease Control and Prevention 2019). Carbapenem resistance has been attributed to the presence of β-lactamases, particularly class B carbapenemases, and metallo-β-lactamases (MBLs) that utilize zinc ions to hydrolyze β-lactam antibiotics (Mourabiti et al., 2025).

Hospital wastewater (HWW) is regarded as a reservoir of antibiotic resistance and carbapenem-resistant bacteria (CRB), including CRPA and CRAB, as well as ARGs and CRGs that contribute to the spread of antibiotic resistance determinants in the clinical setting (Męcik et al., 2024). The chemical and biological composition of HWW differs from that of other types of wastewaters. The composition of HWW is influenced by the type and size of hospitals (number and type of hospital wards/units), ancillary services (kitchen, laundry, air-conditioning), hospital management policy, and awareness levels of healthcare personnel (Carraro et al., 2018; Kumari et al., 2020). Due to high concentrations of pharmaceuticals, hormones, contrast agents, other medical substances, and antibiotic residues that are excreted with urine and feces, HWW constitutes a more biochemically diverse and demanding environment than typical municipal wastewater (Rozman et al., 2020; Sib et al., 2019). In addition to chemical contaminants, HWW can also contain substantial amounts of pathogenic and potentially pathogenic bacteria, such as *Escherichia coli*, *Pseudomonas aeruginosa*, *Klebsiella pneumoniae*, and *Enterococcus faecium*, which are often detected in wastewater discharged from hospitals that treat contagious diseases and act as sources of many ARGs (Ma et al., 2024). There is considerable evidence to suggest that the presence of ARGs in HWW is linked to antimicrobial consumption and the resistance of clinical isolates in a given hospital. Research has also shown that the prevalence of ARB and ARGs varies depending on the type of antimicrobial drug and bacterial family/genus (Ma et al., 2024; Perry et al., 2021). Legal regulations concerning HWW treatment methods and the qualitative requirements for treated HWW differ across countries. In many developing nations, untreated HWW is evacuated directly to public sewers and/or surface waters, whereas in other regions of the world, HWW is classified as industrial wastewater that has to undergo preliminary treatment before it reaches a municipal wastewater treatment plant (WWTP) or as municipal water that is evacuated directly to public sewers and treated together with other streams of municipal wastewater. In Poland, hospital wastewater is not subject to mandatory disinfection before being discharged into the municipal sewage system. This is only required for hospital wastewater from hospitals with infectious disease wards (Męcik et al., 2024). However, these solutions effectively eliminate conventional contaminants, but they do not guarantee complete removal of micropollutants from HWW and WWTPS (Werkneh and Islam, 2023). Therefore, additional disinfection methods, such as chlorination, ozonation and UV treatment, should be applied before HWW is evacuated from a hospital or a WWTP to minimize the spread of dangerous wastewater components, including ARGs that contribute to the dissemination of antibiotic resistance among environmental microorganisms (Chen et al., 2014; Ramírez-Coronel et al., 2024; Stefaniak et al., 2025).

In view of the above, the aim of this study was to assess the prevalence of CRB in HWW discharged (without pretreatment) to the public sewer system and municipal WWTPs by 64 healthcare facilities in all Polish regions and to determine the relationships between CRB counts, sampling season, and the location of HWW sampling sites. Special attention was paid to priority pathogens in the HWW microbiome, particularly cream-colored colonies (in this study called CCRC – Cream Carbapenem Resistant Colonies), which could be CRAB and CRPA that pose a significant threat to public health and the environment. To our the best knowledge, this is the first nationwide, large-scale investigation in Poland addressing carbapenem resistance in *Pseudomonas aeruginosa* and *Acinetobacter baumannii* in hospital wastewater. In contrast to previous studies focusing on local prevalence, this paper introduces a comprehensive surveillance framework that integrates quantitative antibiotic residue analysis with high-resolution genotypic fingerprinting across diverse clinical catchments. By positioning HWW as a dynamic 'biochemical reactor' rather than a passive waste stream, we provide a conceptual advancement in understanding how sub-inhibitory concentrations of novel enzyme inhibitors (e.g., vaborbactam) modulate the environmental resistome, aligning with the global 'One Health' priorities for standardized AMR monitoring.

## 2. Materials and Methods

### 2.1. Sampling Sites

Samples of HWW were collected from 64 hospitals located in all 16 Polish voivodeships (3 to 5 hospitals per voivodeship). Polish voivodeships (Figure S1) differ considerably in population, population density, and socio-economic development. The voivodeships located in central (Mazovia) and south-western (Silesia and Wielkopolska) Poland are most populous, whereas the most sparsely populated voivodeships are found in south-western (Opole) and western (Lubusz) parts of the country. Population density is highest in Silesia (approx. 350 persons/km$^2$) and lowest in north-eastern Poland (Podlasie and Warmia-Masuria) (below 60 persons/km$^2$) (Statistics Poland 2023a, 2024). Mazovia is characterized by the highest GDP per capita, whereas the lowest values of this parameter are noted in eastern Poland (Turczak 2017; Statistics Poland 2022). Polish voivodeships also differ in the urbanization rate which is highest in the voivodeships of Silesia, Mazovia, and Lower Silesia, and lowest in the voivodeships of Opole, Podkarpacie, and Podlasie (Statistics Poland 2023b). These variations affect the availability of public infrastructure and public services, including healthcare facilities, water supply and public sewer systems (Statistics Poland 2023a, 2024).

### 2.2. Hospital Wastewater Sampling

Samples of HWW were collected in the winter and summer of 2024 to analyze the relationship between the prevalence of CRB and carbapenem consumption in Polish hospitals in these seasons. The main aim of the study was to determine the prevalence of pathogenic strains of CRAB and CRPA in the collected HWW samples. Wastewater was sampled in the terminal part of the hospital's sewage installation (the last wastewater manhole that evacuates wastewater to the public sewer, where HWW is mixed with other wastewater streams and directed to the WWTP), which collects wastewater from the entire hospital. Samples for microbial and physicochemical analyses (concentrations of antibiotics and biogenic elements) were collected into two sterile containers with a total volume of 6 L. The studied hospitals differed in size, number and type of wards, bed capacity, and the type and consumption of carbapenem antibiotics. The collected samples consisted of non-disinfected HWW. A total of 128 HWW samples were collected during the study, including 64 in winter and 64 in summer. The samples were transported to a laboratory under refrigerated conditions and analyzed within 24 h after delivery.

### 2.3. Antibiotic Concentrations and Chemical Parameters

Physicochemical analyses were conducted in 5 L of HWW from each of the collected samples; HWW samples were stored at a temperature of -20°C until the determination of carbapenem antibiotics, ions, and physicochemical parameters.

#### 2.3.1. Determination of Carbapenem Antibiotics in HWW Samples

The vaborbactam (VBR) reference standard was obtained from the Cayman Chemical Company (Ann Arbor, MI, USA), whereas imipenem (IMP), meropenem (MEM), and cilastatin (CIL) were purchased from Merck/MilliporeSigma (Burlington, MA, USA). Acetonitrile (LC-MS grade) and formic acid (FA) (≥99%) were supplied by Avantor (Radnor, PA, USA). Pure methanol (MeOH) and hydrochloric acid (35–38%; HCl) were acquired from Chempur (Ruda Śląska, Poland). 4-morpholinepropanesulfonic acid (MOPS) was obtained from Merck/MilliporeSigma (Burlington, MA, USA). High-purity nitrogen (5.0 grade) was supplied by SIAD (Ruda Śląska, Poland). Standard solutions of the analytes were prepared at a concentration of 1 mg/mL. CIL and VBR standards were dissolved in MeOH, whereas IMP and MEM standards were prepared using a MOPS buffer according to the method described by Zou et al. (2019). Standard stock solutions were stored in a freezer at -20 °C. Before pretreatment, all wastewater samples were centrifuged and passed through 0.7 μm MN-GF-1 glass fiber filters (MACHEREY-NAGEL, Düren, Germany). Antibiotics were extracted using a previously developed SPE procedure. In the first step, 1 L of wastewater was adjusted to a pH of 2.5 with concentrated HCl. Wastewater samples were passed through Oasis HLB (500 mg, 6 mL) cartridges conditioned with 6 mL of MeOH, 6 mL of 0.1 M HCl, and 6 mL of distilled water. In the last step, wastewater samples were eluted with 6 mL of MeOH and 6 mL of 0.1% formic acid in water. The extracts were evaporated to 1 mL under a nitrogen stream with a MultiVap 10 system (LabTech, Sorisole, Italy), filtered with a PES syringe filter, and analyzed by LC-MS/MS. Each sample was prepared in triplicate, and each preparation was analyzed three times. Chromatographic separation was performed in an UltiMate 3000 RS Dionex HPLC system (Dionex Corporation, Sunnyvale, CA, USA). The ZORBAX RRHD Eclipse Plus C18 (2.1×50 mm, 1.8 μm) column was the stationary phase, whereas 0.3% formic acid in water (A) or methanol (B) was the mobile phase. Two gradients were applied in the analysis of MEM and IMP (positive ionization, ESI(+)) and VBR and CIL (negative ionization, ESI(–)). In positive ionization mode, the gradient began with 90% A and 10% B. At 1.0 min, the ratio was shifted to 60% A and 40% B and was maintained until 2.0 min. At 6.0 min, gradient composition was 50% A and 50% B. In negative ionization mode, the gradient began with a ratio of 90% A and 10% B, shifting to 30% A and 70% B at 2.0 min. At 3.5 min, the mobile phase was adjusted to 0% A and 100% B and was maintained until 4.5 min. The flow rate (0.3 mL/min), injection volume (5 μL), and temperature inside the column (30°C) were kept constant during the analysis. The LC detector was a QTRAP 4000 hybrid triple quadrupole-linear ion trap tandem mass spectrometer (AB Sciex LLC, Framingham, MA, USA) equipped with a Turbo V™ ionization source operating in electrospray ionization (ESI) mode. Antibiotics were detected in multiple reaction monitoring (MRM) mode, with two MRM transitions scheduled for each compound (Table S1). The dwell time of each transition reaction was set at 200 ms. The ESI(+) parameters were optimized as follows: curtain gas pressure (CUR) – 40 psi, nebulizing gas pressure (GS1) – 60 psi, drying gas pressure (GS2) – 50 psi, ion source temperature (TEM) – 550°C, and ion spray voltage (IS) at – 5500 V. The following settings were applied in ESI(–): CUR – 50 psi, GS1 – 60 psi, GS2 – 70 psi, TEM – 550°C, and IS – -4500 V. High collision gas (CAD) settings were applied in both ionization modes. The entrance potential (EP) was set at 7 V for ESI(+) and at -7 V for ESI(–). Data were acquired using Analyst 1.6 software.

The method was validated using a wastewater matrix (blank samples, free of IMP, MEM, VBR, and CIL) containing working solutions of the analyzed antibiotics with known concentrations. During the analysis, IMP was unstable in wastewater samples, and it degraded rapidly (in less than 1 hour) after addition. Imipenem was stable only in the MOPS buffer (pH=7). Similar observations were made by Khurana et al. (2024). Therefore, the method could not be validated and IMP could not be determined in real-world samples. The calibration curve for the analytes ranged from 0.5 ng/L to 500 ng/L for MEM and CIL, and from 5.0 to 500 ng/L for VBR. The $r^2$ values for the obtained calibration curves were determined at ≥0.994. Precision, accuracy, matrix effect (ME), and recovery were analyzed at a concentration of 10, 250, and 400 ng/L. Precision, expressed by the coefficient of variation (CV), ranged from 0.2% to 1.81%, whereas accuracy, expressed by the relative error (RE), ranged from -13.3% to 13.1%. The ME ranged from –8.43% to 11.6%. The recovery of the examined pharmaceuticals ranged from 50.7% to 97.9%.

### 2.3.2. Physicochemical Parameters of HWW Samples

The following components were quantified in HWW samples: total organic carbon (TOC), total nitrogen (TN), ammonia nitrogen ($N\text{-}NH_4^+$), anions that play an important role in environmental analyses, i.e. $F^-$, $Cl^-$, $NO_2^-$ (subsequently recalculated to $N\text{-}NO_2^-$), $NO_3^-$ (subsequently recalculated to $N\text{-}NO_3$), $Br^-$, $PO_4^{3-}$, $SO_4^{2-}$, and organic nitrogen ($N_{org}$). Total organic carbon and TC were determined with a TOC-LCPH Total Organic Carbon Analyzer (Shimadzu, Japan) equipped with a TNM-L Total Nitrogen Unit (Shimadzu, Japan) that supports the determination of both TOC and TN by the combustion catalytic oxidation technique. The device was equipped with two detectors: a non-dispersive infrared detector (NDIR) and a chemiluminescent detector for determining carbon and nitrogen compounds, respectively. The samples were combusted at a temperature of 720 °C on a platinum catalyst. High-purity air was used as the carrier gas (flow rate = 150 mL/min). Total organic carbon was determined as the difference between the concentration of total carbon (TC) and inorganic carbon (IC). The measurements were performed, and the results were analyzed using LabSolutionTM TOC PC software. Ammonia nitrogen ($N\text{-}NH_4^+$) was determined photometrically using Spectroquant® Ammonium Reagent Tests (Merck Millipore, Poland) and a NOVA 60 A Spectroquant® spectrophotometer based on a method that is analogous to EPA 350.1, APHA 4500-$NH_3$ F, ISO 7150-1, and DIN 38406-5 standards. The presence of selected anions ($F^-$, $Cl^-$, $NO_2^-$, $NO_3^-$, $Br^-$, $PO_4^{3-}$, and $SO_4^{2-}$) in the samples was determined using a Thermo Scientific Dionex Aquion ion chromatograph with a built-in isocratic pump, a vacuum degassing system, and a thermostatted conductivity measuring cell (A.G.A. Analytical, Poland). The ion chromatograph was equipped with a Thermo Scientific Dionex AS-DV autosampler that supports automatic sample feeding and filtration (A.G.A. Analytical, Poland). An IonPac AS9-HC analytical column (4 x 250 mm) with an IonPac AG9-HC pre-column (4 x 50 mm) was used as the stationary phase, and the eluent solution for the IonPac

AS9-HC column was used as the mobile phase (flow rate = 1 mL/min). The concentrations of selected ions were calculated using a 5-point calibration curve that was created and validated with Thermo Scientific Dionex Seven Anion Standards for ion chromatography (A.G.A. Analytical, Poland). The limit of quantification for the tested anions (LOQ) was set as the first lowest point on the specific calibration curve, and the analytes were identified based on their retention time. The relevant information is presented in Table S2. Based on their molar mass, the determined concentrations of $NO_2^-$ and $NO_3^-$ were recalculated to direct concentrations of nitrite nitrogen (N-$NO_2^-$) and nitrate nitrogen (N-$NO_3^-$). Thermo Scientific Dionex Chromeleon 7.2 software was used to control the chromatograph and collect and process analytical data. The content of $N_{org}$ in the samples was calculated by subtracting the concentrations of other nitrogen species (N-$NH_4$+, N-$NO_2^-$ and N-$NO_3^-$) from TN content.

### 2.4. Isolation of CR Bacteria

Before starting the tests on the relevant samples, pilot tests were carried out to select the best medium for isolating CRAB and CRPA. Serial dilutions of HWW samples were performed to isolate *Acinetobacter* sp. and *Pseudomonas* sp. Sample dilutions with a concentration of $10^{-1}$ to $10^{-5}$ HWW, depending on the visual assessment of the samples and probable contamination levels, were cultured on CHROMagar mSuperCARBA medium (GRASO). The medium used for microbiological analyses supports the identification of bacteria producing NDM, KPC, VIM, IMP, and OXA carbapenemases based on differences in the color of the emerged colonies. This medium promotes only the growth of CR colonies, which consequently means a smaller population of the strains analyzed. Therefore, individual, exclusively carbapenem-resistant colonies of a creamy color and/or transparent, classified by the manufacturer as potential pathogens of the genera *Acinetobacter* and *Pseudomonas*, were selected for further analysis. The emerged cream-colored and/or translucent strains were counted and compared with the total number of strains cultured on the CHROMagar mSuperCARBA medium. Whenever possible, approximately 20 isolates from each sample were selected for further analysis. The isolates were passaged onto Tryptone Soya Agar (Sigma-Aldrich, Merck) to obtain pure colonies. The isolates cultured on both media were incubated at a temperature of 37°C (Memmert INP 200 Incubator) for 24 h. The obtained strains were passaged onto 2 mL of liquid LB medium (Sigma-Aldrich, Merck) with the addition of 10% glycerol and stored at a temperature of -20°C until further analysis.

### 2.5. Genomic DNA Extraction

Bacterial genomic DNA was extracted by heat lysis (Dashti et al., 2009) with some modifications (Osińska et al., 2017). Frozen strains were once again passaged onto Tryptone Soya Agar (Sigma-Aldrich, Merck) and incubated at a temperature of 37°C (Memmert INP 200 Incubator) for 24 h. Each emerged strain was suspended in 1 mL of phosphate-buffered saline (PBS) solution (1.5 mL 1xPBS), and the bacterial suspension was centrifuged for several seconds at full speed to collect the pellet. The supernatant was removed. The pellet was combined with 0.5 mL of double distilled water (dd$H_2O$) and boiled at a temperature of 95°C for 10 minutes in a Grant QBD4 Block Heater (Grant Instruments). After thermal processing, the samples were centrifuged at 5000 rpm and 4°C for 5 min. The supernatant (containing DNA) was transferred to a sterile Eppendorf tube. Genomic DNA was extracted from a total of 2450 bacterial isolates that were obtained from HWW samples collected in both analytical seasons. The quality and quantity of genetic material was assessed with a Multiskan SkyMicroplate Spectrophotometer (Thermo Scientific, Waltham, MA, USA). DNA was stored at a temperature of -20°C until further analysis.

## 2.6. Identification of CRAB and CRPA

The isolated bacterial strains were identified as *A. baumannii* and *P. aeruginosa* based on the results of a standard polymerase chain reaction (PCR). A multiplex PCR assay was designed to target the *opr*L gene in *P. aeruginosa* (De Vos et al., 1997) (with some modifications proposed by Al-Ahmadi and Roodsari (2016)) and the *Ab*-ITS gene in *A. baumannii* (Chen et al., 2007) (with some modifications proposed by Hubeny et al. (2022)). The reaction mix with a total volume of 15 μL was composed of: NZYTaq II 2xGreen Master Mix (NZYTECH, Portugal), 10 μM of the corresponding primer pairs, and 1 μL of matrix DNA (Table S3). The reaction consisted of the following steps: initial denaturation at 94°C for 5 min, followed by 30 cycles at 94°C for 1 min, 56°C for 1 min, and 72°C for 1 min, and a final extension step at 72°C for 10 min in Mastercycler Nexus X2 and Vapo Protect Mastercycler Pro (Eppendorf, Germany). In each reaction, negative and positive controls were analyzed simultaneously. The reference strains of *P. aeruginosa* ATCC 9027 and *A. baumannii* PCM 2740 were the positive controls, whereas double distilled water (dd$H_2O$) was the negative control. The reaction products were verified by electrophoresis on 1% agarose gel with the addition of 0.5 mg/mL of ethidium bromide (BioRad PowerPac Basic Power Supply). Electrophoresis gels were visualized with Gel Doc EZ and Image Lab software (Bio-Rad Laboratories, CA, USA). The randomly selected isolates were identified by 16S rRNA gene sequencing. Universal bacteria primers, 27F and 1492R, were used to amplify nearly fulllength 16S rRNA genes in accordance with a previously described method (Gillan et al., 1998).

## 2.7. Identification of Clones by ERIC-PCR

The presence of clonal similarities between the isolated *A. baumannii* and *P. aeruginosa* strains was determined by enterobacterial repetitive intergenic consensus (ERIC)-PCR based on the method described by Versalovic (Versalovic et al., 1991) with some modifications (Osińska et al., 2017; Rolbiecki et al., 2021). The reaction involved ERIC1 and ERIC2 primers (Table S4), and the products were separated by electrophoresis on 1.5% agarose gel with the addition of 0.5 mg/mL of ethidium bromide. Electrophoresis gels were visualized with Gel Doc EZ and Image Lab software (Bio-Rad Laboratories, CA, USA). *Acinetobacter baumannii* and *P. aeruginosa* strains were identified using GelJ software (Heras et al., 2015) that facilitates automatic comparison of DNA fingerprint images on the same gel and across gels. The images were analyzed automatically and corrected manually if necessary. For the needs of a detailed analysis, a genetic similarity matrix was developed based on the values of the Dice coefficient and Pearson's correlation coefficient. The phylogenetic tree was built with the Unweighted Pair Group Method with Arithmetic Mean (UPGMA) algorithm. A 5% error margin was applied to compensate for variability in electrophoretic separation and to normalize gel images.

### 2.8. Detection of ARGs and CRGs by PCR

A total of 14 AB strains and 242 PA strains (17 PA strains were ultimately identified as genetic clones and were excluded from the analysis) were subjected to further analysis in a series of PCR assays. *Acinetobacter baumannii* strains were analyzed for CRGs (*bla*$_{\text{OXA23-like}}$, *bla*$_{\text{OXA24-like}}$, *bla*$_{\text{OXA51-like}}$, *bla*$_{\text{OXA58-like}}$, *bla*$_{\text{BIC}}$, *bla*$_{\text{SPM}}$, *bla*$_{\text{AIM}}$, and *bla*$_{\text{GIM}}$) (Amudhan et al., 2011; Poirel et al., 2011), genes encoding the production of extended-spectrum beta-lactamases (ESBLs) that confer resistance to beta-lactam antibiotics (*bla*$_{\text{OXA}}$, *bla*$_{\text{CTX}}$, *bla*$_{\text{TEM}}$, *bla*$_{\text{SHV}}$) (Kim et al., 2009), tetracycline resistance genes (*tet*A and *tet*M) (Ng et al., 2001), sulfonamide resistance genes (*sul*1 and *sul*2) (Pei et al., 2006), macrolide resistance genes (*qep*A) (Li et al., 2012; Osińska et al., 2016), genes encoding resistance to macrolide-lincosamide-streptogramin antibiotics (*erm*F) (Eitel et al., 2013), and genes encoding the production of MBLs (*bla*$_{\text{VIM}}$, *bla*$_{\text{NDM}}$ and *bla*$_{\text{IMP}}$) (Amudhan et al., 2011; Goudarzi et al., 2015). Due to the limited efficacy of primers targeting the *bla*$_{\text{VIM}}$ gene in the analyzed bacteria, primers with a length of 500 bp were used in *P. aeruginosa* (Amudhan et al., 2011) and primers with a length of 247 bp were used in *A. baumannii* (Goudarzi et al., 2015). The analyses of PA strains did not target *bla*$_{\text{OXA23-like}}$, *bla*$_{\text{OXA24-like}}$, *bla*$_{\text{OXA51-like}}$, and *bla*$_{\text{OXA58-like}}$ genes due to the scarcity of published information about these genes in *P. aeruginosa*. PCR and multiplex PCR assays were conducted using the NZYTaq II 2xGreen Master Mix (NZYTECH, Portugal), 10 μM of the corresponding primer pairs, and 1 μL of matrix DNA, based on the manufacturer's protocol (Table S4). The required controls were applied in all assays. The reaction products were analyzed by electrophoresis on 1% agarose gel with the addition of 0.5 mg/mL of ethidium bromide. Electrophoresis gels were visualized with Gel Doc EZ and Image Lab software (Bio-Rad Laboratories, CA, USA).

### 2.9. Statistical Analysis

Descriptive statistical analyses were conducted in MS Excel (Microsoft Corporation). Detailed and inferential statistical analyses were performed in SPSS Statistics 29.0.2. (IBM Inc.). Data normality was assessed using the Shapiro–Wilk test. As the majority of variables deviated from normality (o$p < 0.05$), non-parametric statistical tests were applied. The threshold for statistical significance was set at $\alpha = 0.05$. Differences in the counts of isolated carbapenem-resistant colonies across sampling seasons and sites were assessed using the Kruskal–Wallis test, followed by the Mann–Whitney U test (two-group comparisons), and the Friedman test for repeated measurements. Analyses were stratified according to sampling season and sampling site. Associations between hospital carbapenem consumption, carbapenem-resistant bacteria (CRB) counts, and physicochemical parameters of hospital wastewater (HWW) were assessed using Spearman's rank correlation. Principal Component Analysis (PCA) was performed in Statistica v.13.3 (TIBCO Software Inc.) using standardized variables. PCA included hospital-related variables (bed capacity, daily wastewater volume, annual carbapenem consumption), microbiological variables (CCRC and ACRC counts), antibiotic concentrations in HWW, and physicochemical parameters of HWW.

## 3. Results

### 3.1. Antibiotic Concentrations and Chemical Parameters of the Studied HWW

#### 3.1.1. Carbapenem Concentrations

The analysis of the concentrations of selected carbapenem antibiotics in HWW samples collected from hospitals in all Polish voivodeships revealed differences across seasons and regions. Median and standard deviation values (Table S5) point to considerable variation in the concentrations of the studied antimicrobials between voivodeships and sampling seasons. The analyses conducted showed significant variability in antibiotic concentrations depending on the geographical region and research season. In winter, the presence of cilastatin/imipenem was detected in HWW samples acquired in all regions (Figure 1 A and C), and the highest concentrations of these drugs in HWW samples were noted in north-eastern and central Poland in the range from 0.00 to 2375.24 ng/L and from 23.16 to 17315.79 ng/L, respectively, indicating intensified hospital antibiotic use in these regions. Cilastatin/imipenem levels were the lowest in HWW samples from north-central Poland, which may suggest regional differences in prescribing practices. Meropenem and vaborbactam displayed more region-specific distribution patterns, with localized concentration peaks rather than uniform nationwide occurrence. Meropenem was the second most frequently detected antibiotic in HWW, and its concentration peaked in samples collected from hospitals in south-western Poland. In this region, meropenem concentration in HWW samples ranged from 0.00 to 163.23 ng/L. In HWW samples from other regions, meropenem concentration was very low and did not exceed 0.00 – 23.85 ng/L or was below the method detection limit. Vaborbactam levels peaked in HWW samples from south-western and northern regions ranging from 40.41 to 62.75 ng/L and from 0.00 to 85.13 ng/L, respectively. In HWW samples collected in the remaining regions, the concentration of vaborbactam did not exceed 49.77 ng/L or this antibiotic was not detected.

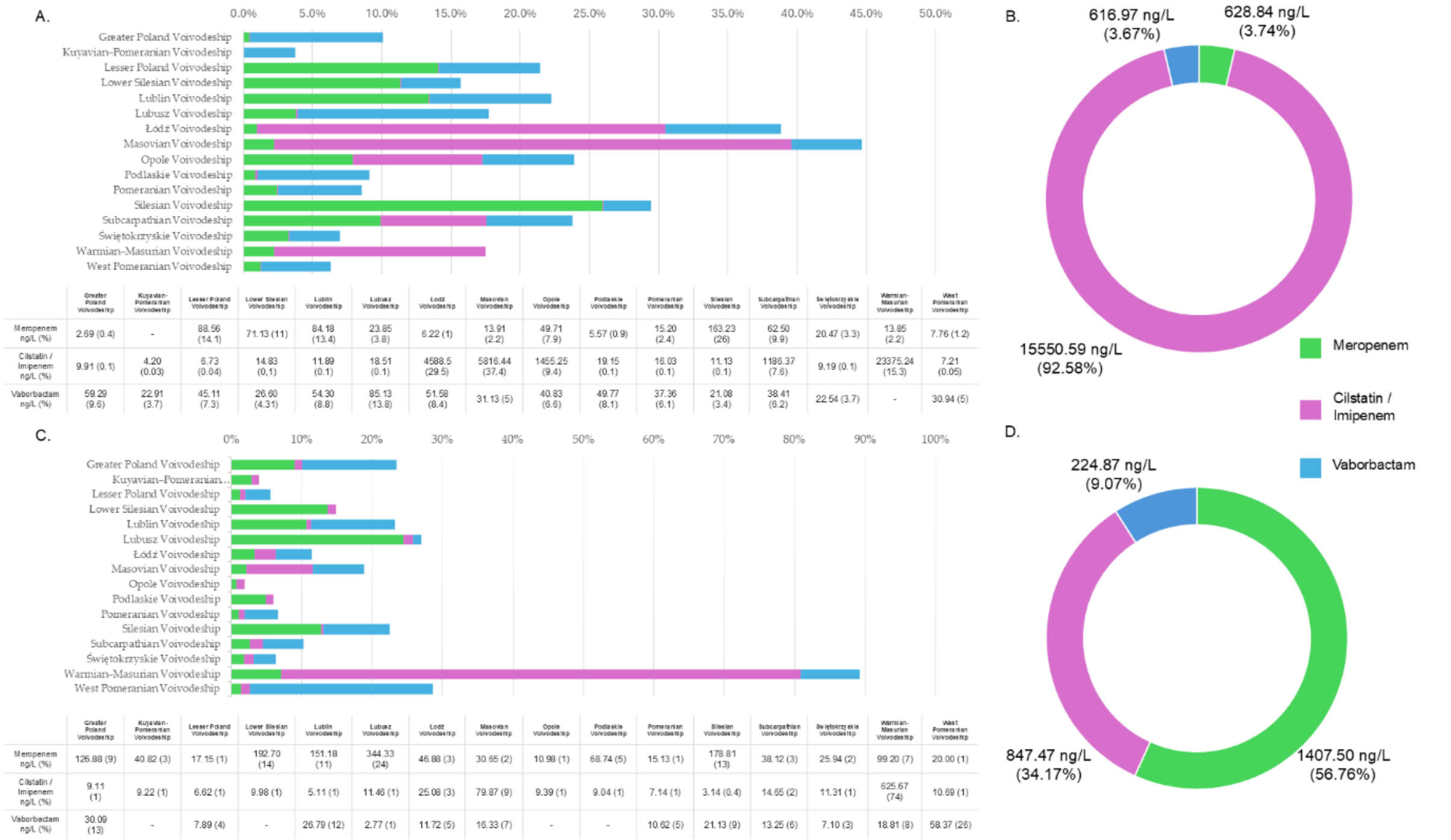

**Figure 1.** Concentrations of the tested carbapenem antibiotics (meropenem, cilastatin/imipenem, vaborbactam) in samples of hospital wastewater (HWW) collected in each voivodeship in winter (A) and summer (C), expressed as a percentage of the average nationwide concentrations of antibiotics in each analyzed season.

The average nationwide concentrations of the tested antibiotics in HWW in winter (B) and summer (D) are also presented.

In summer (Figure 1 B and D) carbapenem concentrations in HWW samples were lower and more heterogeneous compared with winter samples. Cilastatin/imipenem was found in the highest average concentration in HWW samples from north-eastern and central Poland, which reached 6.15 – 2180.12 ng/L and 2.85 - 294.37 ng/L, respectively. In the remaining regions, cilastatin/imipenem levels in HWW samples were considerably low and did not exceed 13.00 – 51.98 ng/L. Meropenem emerged as the most frequently detected antibiotic, with elevated concentrations in HWW samples from western and eastern Poland, ranging from 9.35 to 428.15 ng/L and from 1.60 to 950.80 ng/L, respectively, indicating sustained use during this period. In HWW samples from the remaining regions, meropenem levels were determined in the range of 6.33 – 171.08 ng/L. Vaborbactam was the least prevalent carbapenem in HWW samples collected in summer, with low concentrations or non-detection in most regions. The highest average concentrations were observed in HWW samples from eastern and western Poland, ranging from 11.29 to 48.93 ng/L and from 12.13 to 104.96 ng/L, respectively. In the remaining regions, vaborbactam levels were very low or the drug was not detected in HWW samples.

### 3.1.2. Physicochemical parameters

The physicochemical parameters of HWW samples differed across seasons and regions (Table S6). In winter, TOC was determined in the range of 71.35 ± 52.94 mg/L to 288.80 ± 184.15 mg/L, and the highest values of this parameter were noted in HWW samples from western and central Poland. Total nitrogen and $N\text{-}NH_4^+$ levels in the examined HWW samples ranged from 9.08 ± 3.09 mg/L to 21.96 ± 22.18 mg/L and from 2.99 ± 8.62 mg/L to 20.08 ± 22.73 mg/L, respectively, and the values of both parameters were fairly similar in all regions. The content of $N_{org}$ was highest in HWW samples from central and north-central Poland, and the value of this parameter ranged from 0.00 ± 0.13 mg/L to 9.68 ± 19.82 mg/L in all winter samples. Most HWW samples had a similar pH of 6.50 ± 0.00 to 7.25 ± 0.95, and the highest pH was noted in samples from north-eastern, central, and southern Poland. In the group of the analyzed ions ($F^-$, $Cl^-$, $N\text{-}NO_2^-$, $Br^-$. $N\text{-}NO_3^-$, $PO_4^{3-}$, $SO_4^{2-}$), chloride and sulfide ions were detected most frequently, within a concentration range of 61.34 ± 38.28 mg/L to 396.99 ± 340.19 mg/L and 39.12 ± 12.86 mg/L to 75.31 ± 140.41 mg/L, respectively. The concentrations of the remaining ions in all HWW samples did not exceed 9.19 ± 20.64 mg/L. In summer, TOC levels in HWW samples ranged from 78.93 ± 111.82 mg/L to 169.78 ± 368.43 mg/L, and the highest values were observed in samples from western and central Poland. In turn, the content of TN and $N\text{-}NH_4^+$ ions was fairly similar in HWW samples from all regions, and these parameters peaked at 33.87 ± 13.85 mg/L for TN and 30.55 ± 13.96 mg/L for $N\text{-}NH_4^+$ ions in HWW samples from south-western Poland. Samples from south-central Poland were characterized by the highest concentration of $N_{org}$, ranging from 0.00 ± 0.16 mg/L to 3.42 ± 5.36 mg/L. In some HWW samples, $N_{org}$ levels were below the method detection limit. The pH of the studied HWW samples was relatively stable and remained in the range of 6.95 ± 1.46 to 7.00 ± 0.56. The highest pH was noted in HWW samples from eastern and northern Poland. Similarly to winter samples, an analysis of ion concentrations in summer samples ($F^-$, $Cl^-$, $N\text{-}NO_2^-$, $Br^-$. $N\text{-}NO_3^-$, $PO_4^{3-}$, $SO_4^{2-}$) revealed that chloride and sulfide ions were detected most frequently. The concentration of chloride ions ranged from 75.65 ± 25.17 mg/L to 593.21 ± 273.47 mg/L, and the concentration of sulfide ions ranged from 0.00 ± 12.17 mg/L do 101.54 ± 86.35 mg/L. Summer

HWW samples were also significantly more abundant in fluoride and phosphate ions. The highest concentration of fluoride ions was determined in HWW samples from western Poland (0.41 ± 57.60 mg/L), whereas the highest concentration of phosphate ions was observed in HWW samples from south-central Poland (0.00 ± 58.61 mg/L). The levels of the remaining ions ($N\text{-}NO_2^-$, $Br^-$, $N\text{-}NO_3^-$) in summer samples did not exceed 0.00 ± 5.29 mg/L. The study revealed that the physicochemical parameters of HWW approximated the range of values noted by other researchers in municipal wastewater (Agarwal et al., 2022), which indicates that the examined samples did not raise environmental concerns. However, in addition to standard physicochemical tests, advanced instrumental and microbiological analyses involving molecular biology techniques are required to reliably assess the actual threats posed by HWW.

### 3.2. Nationwide Prevalence of Carbapenem Resistance

In both sampling seasons (winter and summer of 2024), qualitative and quantitative analyses of bacterial colonies cultured on a selective medium revealed considerable differences in CCRC and ACRC counts between seasons and regions (Figure S1, Table S7). In winter, CCRC counts ranged from $3.00 \times 10^3$ to $2.00 \times 10^5$ colony-forming units (CFU)/mL, and the highest values were noted in HWW samples from southern and central Poland (Figure 2A). The average number of CCRC was highest in south-western and central Poland and lowest in north-eastern and east-central Poland. In winter, ACRC counts ranged from $5.00 \times 10^3$ to $8.00 \times 10^5$ CFU/mL and was highest in HWW samples from southern and south-western Poland (Figure 2B). The abundance of these bacterial colonies was highest in HWW samples collected in the voivodeships of Lower Silesia, Opole, and Łódź, and lowest in HWW samples collected in the voivodeships of Mazovia and Warmia-Masuria. Notably, the maximum counts of both CCRC and ACRC in HWW samples collected in summer were an order of magnitude higher than those recorded in winter HWW samples, indicating a significant seasonal escalation in carbapenem-resistant bacteria concentrations across all studied regions.

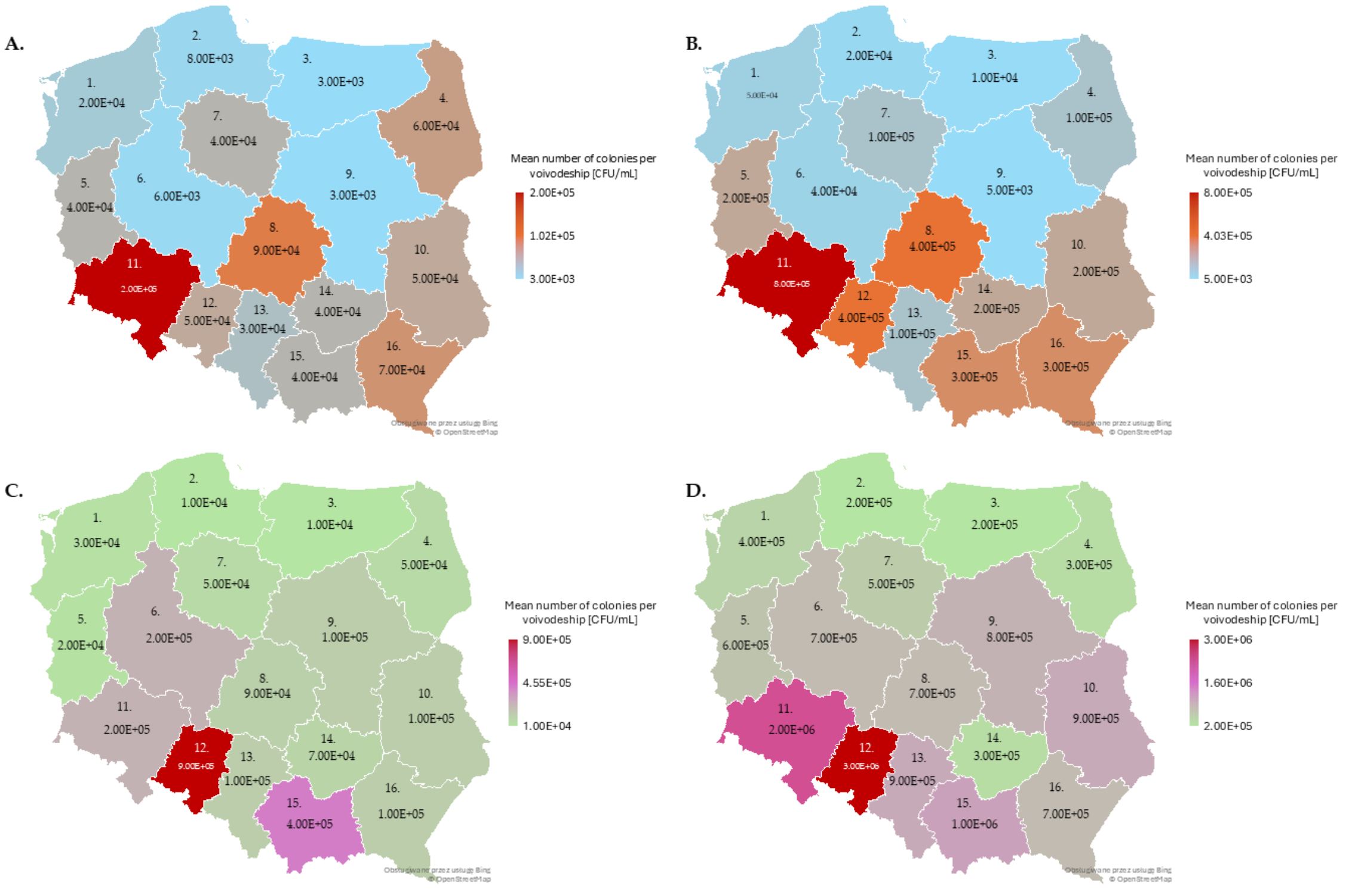

**1. West Pomerania, 2. Pomerania, 3. Warmia-Masuria, 4. Podlasie, 5. Lubusz, 6. Wielkopolska, 7. Kuyavia-Pomerania, 8. Łódź, 9. Mazovia, 10. Lublin, 11. Lower Silesia, 12. Opole, 13. Silesia, 14. Świętokrzyskie, 15. Małopolska, 16. Podkarpacie..**

**Figure 2.** Heatmap of carbapenem-resistant bacteria in Polish voivodeships based on the mean number of colony-forming units (CFU) in 1 mL of HWW. Voivodeships with the lowest percentage of resistant bacteria in winter are marked in light blue (A, B) and those with the lowest percentage of resistant bacteria in summer are marked in light green (C, D). The resistance maps refer to CCRC (A, C) and ACRC (B, D), respectively. Voivodships with an intermediate number of CFU/mL in winter are marked in orange (A, B) and those with an intermediate number of CFU/mL in summer are marked in purple (C, D). Voivodeships with the highest percentage of resistant bacteria in winter (A, B) and summer (C, D) are marked in dark red. Voivodeships are denoted by numbers 1 to 16. Source: own elaboration in MS Office.

The number of CCRC potentially indicative of *A. baumannii* and *P. aeruginosa* was higher in HWW samples collected in summer ($1.00 \times 10^4$ to $9.00 \times 10^5$ CFU/mL) than in those collected in winter, and the highest concentrations of these bacteria were noted in southern and western Polish regions (Figure 2C). The abundance of CCRC was highest in HWW samples from Opole and Małopolska voivodeships, and lowest in HWW samples from northern Poland. The counts of ACRC were also higher in summer ($2.00 \times 10^5$ to $3.00 \times 10^6$ CFU/mL), and the highest values were found in HWW samples from southern and south-western Poland (Figure 2D). In this case, the counts of CRB cultured on a selective medium were highest in HWW samples from Opole and Lower Silesia and lowest in Warmia-Masuria and Pomerania (north-eastern Poland).

### 3.3. Prevalence of CRAB and CRPA

A total of 1229 CCRC were isolated from HWW samples collected in winter and 1221 CCRC were isolated from HWW samples collected in summer in all Polish regions. A total of 2450 CRB strains were subjected to further analysis. Based on the results of the PCR assay targeting the *opr*L gene in *P. aeruginosa* and the *Ab*-ITS gene in *A. baumannii*, 225 strains identified as *P. aeruginosa* and only 14 strains identified as *A. baumannii* were chosen from the pre-selected group of CCRC. The number and concentration of CRAB and CRPA strains in HWW samples differed significantly across Polish regions (Figure S2). The number of *A. baumannii* strains was relatively low and dispersed, and they were most prevalent in HWW samples from eastern and north-central Poland. In the remaining HWW samples, in particular those from western and northern Poland, the number of *A. baumannii* strains was low or they were absent. Carbapenem-resistant *A. baumannii* strains were not detected in HWW samples from 11 voivodeships. In turn, CRPA strains were far more prevalent in HWW samples than CRAB, and their geographic distribution was more varied. The highest number of CRAB and CRPA strains was noted in HWW samples from north-eastern and central Poland, which points to the local accumulation of these pathogens. Hospital wastewater samples from the remaining voivodeships were characterized by more balanced proportions of CRAB and CRPA strains, and the studied pathogens were clearly less prevalent in samples from south-eastern and western Poland.

### 3.4. Prevalence of Antibiotic Resistance Genes and ERIC-PCR assay

The standard PCR analysis revealed seasonal variations in the number of ARGs in *A. baumannii* and *P. aeruginosa* strains isolated from HWW samples. Beta-lactamase genes, in particular $bla_{VIM}$ (500 bp), $bla_{CTX}$, and $bla_{TEM}$, were predominant in CRAB strains (Figure S3.A) in HWW samples collected in both seasons. The strains isolated from winter and summer samples harbored $bla_{VIM}$, $bla_{OXA}$, $bla_{OXA58\text{-}like}$, and $bla_{OXA23\text{-}like}$ genes, but these genes were more abundant in isolates obtained from winter samples. The most common metallo-beta-lactamase gene among CRAB strains was $bla_{NDM}$, accounting for approximately 85%, with a higher prevalence detected in strains collected during the winter season. The prevalence of $bla_{OXA\text{-}51\text{-}like}$, $bla_{OXA24\text{-}like}$, and $bla_{OXA23\text{-}like}$ genes was clearly affected by season, and these genes were more abundant in isolates obtained from winter samples. In CRAB strains, no significant differences in the abundance of *sul*1 and $bla_{VIM}$ genes were observed between isolates obtained in winter and summer. The number of strains containing *sul*2, $bla_{CTX}$, and $bla_{OXA23/24/51\text{-}like}$ genes was considerably higher in HWW samples collected in winter, compared with those collected in summer. The *sul*2 gene was more frequently detected in isolates obtained from winter than in summer samples, which may suggest that the consumption of sulfonamides is higher in the winter season. In turn, $bla_{TEM}$, $bla_{CTX}$, and $bla_{OXA51\text{-}like}$ were the most prevalent ARGs in isolates from summer samples. None of the analyzed CRAB strains harbored $bla_{VIM}$ or *qep*A genes. Carbapenem resistance genes (OXA family, ESBL and MBLs) accounted for around 30% of all ARGs in strains isolated from winter samples, and the proportion of CRGs increased to 40% in strains isolated from summer samples of HWW. Among all CRAB strains, 7 (50%) isolates had multiple resistance genes determining resistance to various types of antibiotics. The ERIC-PCR assay revealed clear genotypic differences across 14 *A. baumannii* strains isolated from HWW samples that were acquired in different regions and seasons. The average value of the clonal similarity index for all isolates was 0.09. The developed dendrogram (Figure 3A) contains five clusters (differently colored branches), and the first three clusters feature closely related strains (2, 6, and 5 strains in each cluster, respectively). The two remaining strains were more genetically distant and formed individual clusters, which suggests that these isolates had a different origin or exhibited high genetic variability. The relatedness between strains was linked to sampling site and season, but it didn't determine the degree of relatedness between strains. Specifically, the heat map analysis confirmed that high genetic similarity (Dice Pearson

> 97%) was predominantly restricted to small intra-cluster groups, while the majority of strains exhibited a high degree of genetic divergence across different geographical locations. The clustering analysis revealed strong relationships between strains isolated from HWW samples from central Poland and those isolated from sample collected in north-eastern and eastern regions. Nearly all analyzed strains harbored genes encoding MBLs and sulfonamide resistance. Carbapenem resistance genes and ESBLs were detected in strains grouped in all clusters. Genes encoding resistance to tetracyclines and macrolides were identified in strains isolated from HWW samples collected in winter in the most highly urbanized or least populated Polish regions. An analysis of the similarity matrix developed based on the values of the Dice coefficient (100%) and Pearson's correlation coefficient (≥97%) revealed up to 50% genetic similarity between the studied *A. baumannii* strains.

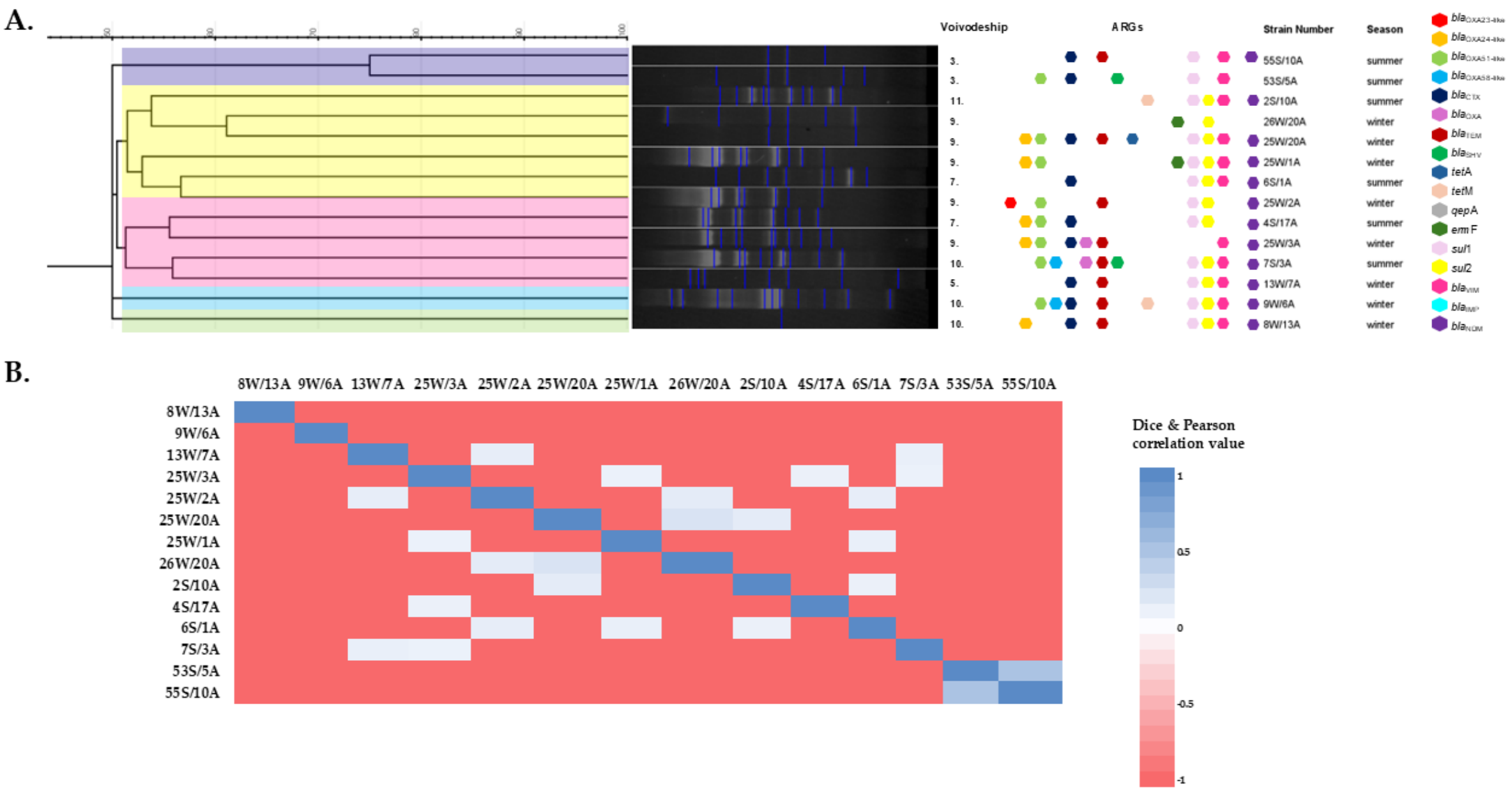

**Figure 3.** The results of ERIC-PCR analysis of CRAB strains presented in a phylogenetic tree (A) and a heat map (B) with additional information on the abundance of ARGs. In the phylogenetic tree (A), clusters of genetically similar strains are marked in different colors, and the length of each branch denotes the degree of genetic similarity relative to the scale. Polish voivodeships are numbered, and differently colored dots denote the presence of specific ARGs. In the heat map (B), high and low levels of genetic similarity between the studied strains are marked in blue and red, respectively, based on the values of the Dice coefficient and Pearson's correlation coefficient. Source: own elaboration in GelJ and MS Office.

The analysis of the prevalence of ARGs in *P. aeruginosa* strains (Figure S3.B.) revealed the presence of the *bla*VIM (247 bp) and *bla*NDM genes encoding MBLs and *bla*CTX and *bla*OXA genes encoding ESBLs in isolates from HWW samples collected in both seasons, but these genes were more abundant in strains obtained from winter samples. *bla*VIM and *bla*CTX genes were more frequently detected in strains isolated from summer samples than from winter samples. The most common metallo-beta-lactamase gene among CRPA strains was also *bla*NDM, accounting for approximately 81% of cases, with a higher prevalence detected in strains collected during the summer season. All CRGs (*bla*GIM, *bla*AIM, *bla*SPM, and *bla*BIC) were identified in isolates selected from winter HWW samples, whereas *bla*GIM and *bla*SPM were also detected in strains obtained from summer samples. In turn, *sul*1 and *sul*2 were more prevalent in isolates from winter samples than in those from summer samples. In addition, HWW samples acquired in winter were characterized by a considerably higher number of strains harboring the *tet*A gene, and a similar trend was noted in the prevalence of the *tet*M gene. In addition to the $bla_{\mathrm{TEM}}$ gene, the isolates from summer samples also carried *tet*A and *sul*1, but these genes were most prevalent in strains isolated from winter samples. Carbapenem resistance genes (MBLs, ESBL, $bla_{\mathrm{GIM}}$, $bla_{\mathrm{AIM}}$, $bla_{\mathrm{SPM}}$, and $bla_{\mathrm{BIC}}$) were identified in 19% of CRPA strains from winter samples and in 38% of CRPA strains from summer samples. The complex topology of the dendrogram, featuring over 30 distinct clusters, illustrates that *P. aeruginosa* populations in Polish HWW maintain a high level of genomic plasticity, with specific resistance profiles like $bla_{\mathrm{NDM}}$ and *sul*1 and *sul*2 appearing consistently across phylogenetically distant lineages. Among the 225 strains isolated from HWW samples collected in both seasons, 57 isolates (25.45%) had multiple resistance genes determining resistance to various types of antibiotics. The ERIC-PCR assay of CRPA strains (Figure 4) revealed considerable genetic heterogeneity within the analyzed population of strains isolated from HWW samples collected in two seasons. An analysis of strains originating from the same sampling site and season revealed 17 clones. Strains belonging to all identified clusters harbored MBL genes (only $bla_{\mathrm{VIM}}$) and sulfonamide resistance genes. Carbapenem resistance genes ($bla_{\mathrm{SPM}}$, $bla_{\mathrm{BIC}}$, $bla_{\mathrm{AIM}}$, and $bla_{\mathrm{GIM}}$) were detected in isolates obtained from summer samples, and their prevalence differed across sampling sites. These genes were predominant in isolates from central regions, characterized by the highest level of economic development, as well as in north-eastern and southern regions, which are far less populated. Genes encoding tetracycline resistance and ESBLs were relatively prevalent and generally accompanied by CRGs, but the former were more frequently identified in isolates from winter samples of HWW.

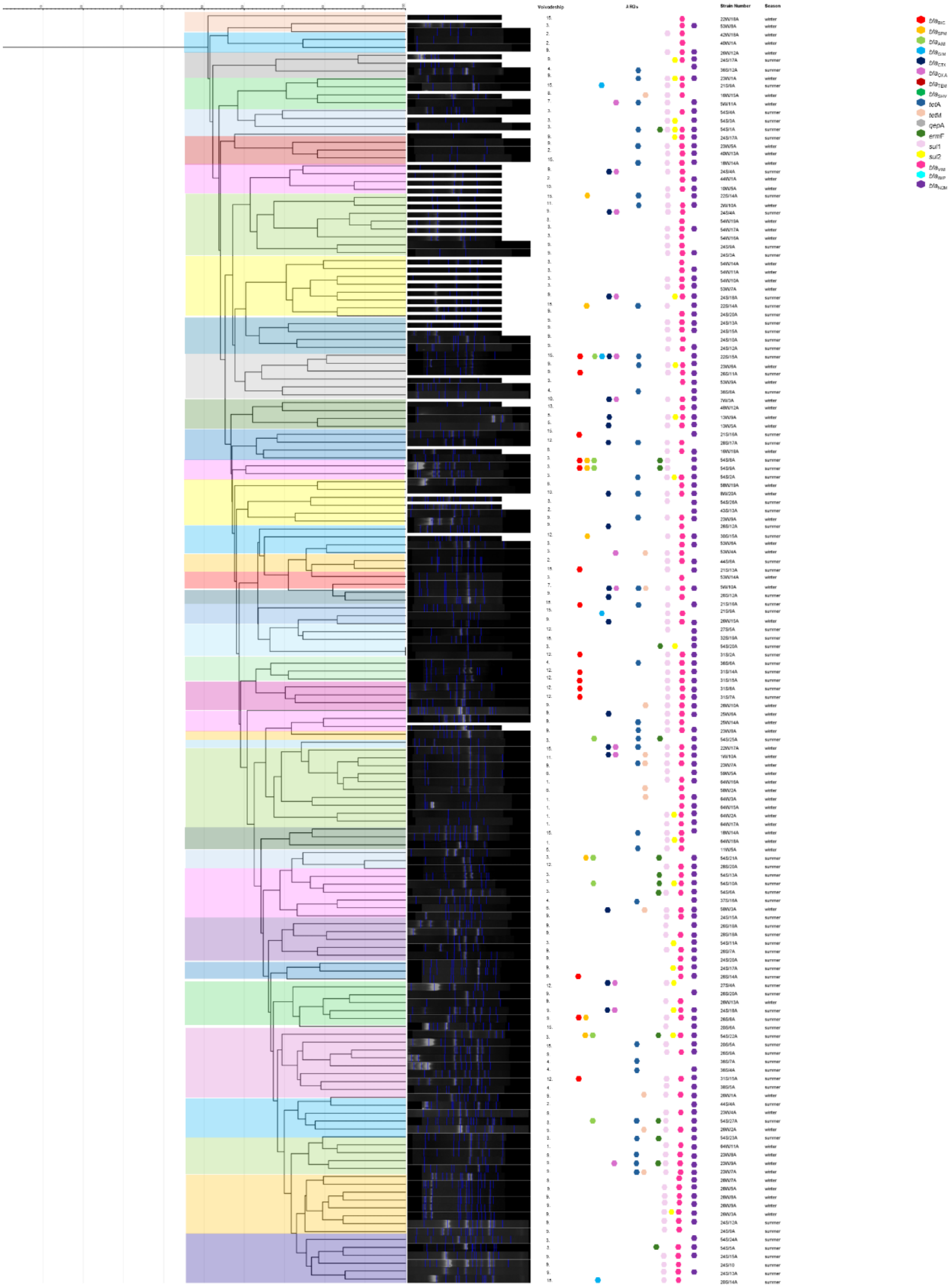

**Figure 4.** The results of ERIC-PCR analysis of CRPA strains presented in a phylogenetic tree. Clusters of genetically similar strains are marked in different colors, and the length of each branch denotes the degree of genetic similarity relative to the scale. Polish voivodeships are numbered, and differently colored dots denote the presence of specific ARGs. Source: own elaboration in GelJ and MS Office.

The phylogenetic tree features more than 30 main clusters grouping isolates with different degrees of genetic similarity. Groups of isolates with 80-95% genetic similarity were identified within clusters. The average value of the clonal similarity index for all isolates was 0.18. The length of dendrogram branches denotes the degree of genetic similarity between isolates and clusters. These findings indicate that the CRPA population in the analyzed HWW samples had a complex genetic structure, featuring both distinct genetic lineages and closely related strains. In addition, the analysis of the similarity matrix (Figure 5) based on the values of correlation coefficients revealed the presence of clones among the strains identified in HWW samples collected in different regions and seasons. The maximum genetic similarity between the remaining *P. aeruginosa* strains was determined at 83%.

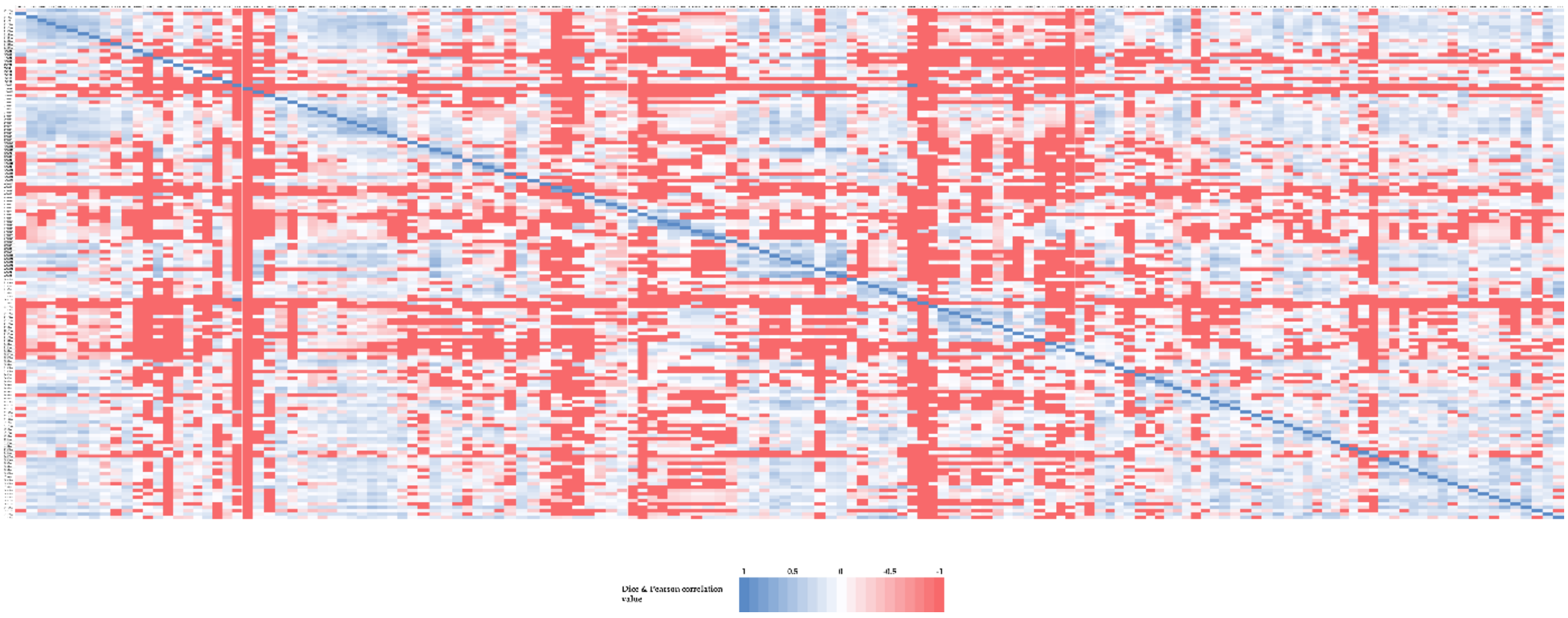

**Figure 5.** The results of ERIC-PCR analysis of CRPA strains presented in a heat map. High and low levels of genetic similarity between the studied strains are marked in blue and red, respectively, based on the values of the Dice coefficient and Pearson's correlation coefficient. Source: own elaboration in GelJ and MS Office.

### 3.5. Statistical Analysis

The presence of relationships between CCRC and ACRC counts in HWW samples collected in winter and summer was determined using the Mann-Whitney U test. The test revealed significant differences in microbial counts between sampling seasons (Figure 6). The analysis of the number of CFUs in the Mann-Whitney U test revealed significant differences in CCRC and ACRC counts between HWW samples collected in summer and winter. Hospital wastewater samples acquired in summer and winter differed significantly in the number of ACRC ($U = 1004$, $p = 0.001$) and CCRC ($U = 1540.00$, $p = 0.015$) CFUs.

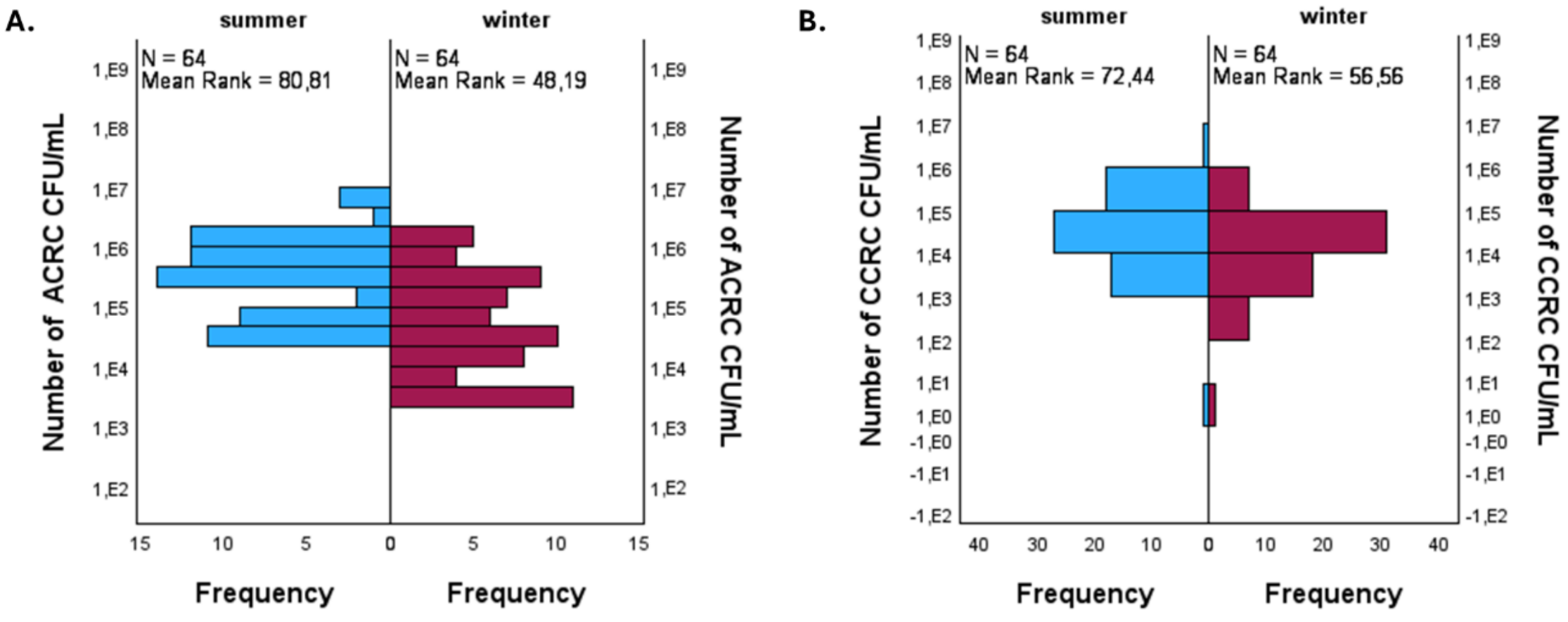

**Figure 6.** The relationships between the counts of all CR colonies (ACRC, CFU/L) (A) and cream-colored CR colonies (CCRC, CFU/L) (B) in HWW samples collected in different seasons determined in the Mann-Whitney U test. The significance level is equal to 0.05. Source: own elaboration in IBM SPSS Statistics and MS Office.

The Kruskal-Wallis test (Figure S4 A) did not reveal significant differences in CRB counts between HWW samples collected in different voivodeships ($p = 0.068$). Significant differences in the abundance of CCRC (Figure S4 B) were noted between HWW samples collected in northern Poland and the remaining regions. It was also found that northern regions, where the prevalence of CCRC in HWW samples was lower, were characterized by smaller hospitals in terms of bed capacity, smaller daily volume of evacuated wastewater, and lower annual consumption of carbapenems. A negative binomial regression analysis revealed that the daily volume of evacuated HWW had a significant negative effect on ACRC and CCRC counts. In turn, carbapenem consumption in hospitals was positively correlated with ARCR counts but had no influence on CCRC counts.

The Friedman test revealed significant differences in the counts of CCRC and ACRC isolated from HWW samples in winter and summer ($\chi^2(1) = 126.00$, $p < 0.001$), which indicates that the abundance of these microorganisms in HWW was significantly affected by season. In addition, the Friedman test also revealed significant differences in selected physicochemical parameters of HWW ($\chi^2(1) = 1167.397$, $p < 0.001$), which suggests that seasonal variations in these parameters were associated with the presence of CRB. The concentrations of the tested antibiotics in HWW samples were not significantly differentiated by season, which indicates that selective pressure exerted by carbapenems was not the main reason for seasonal variations in the size of the CRB population.

Spearman's rank correlation coefficients (Figure 7) were calculated to evaluate the relationship between the annual consumption of carbapenem antibiotics in hospitals (ACCAH), ACRC and CCRC counts in HWW samples, antibiotic concentrations, and the physicochemical parameters of HWW samples. Significant correlations were noted between ACRC and CCRC counts, content of TOC and TN, and vaborbactam concentration in HWW samples. Total CCRC counts were correlated only with vaborbactam concentration and TN content. These results suggest that changes in the total CRB population are linked to TOC content, TN content, and the presence of vaborbactam in HWW samples. The concentrations of the tested antibiotics in HWW samples were mutually correlated, and they were also correlated with ACCAH in the studied hospitals, which confirms that antibiotic consumption in hospitals directly influences antimicrobial levels in HWW.

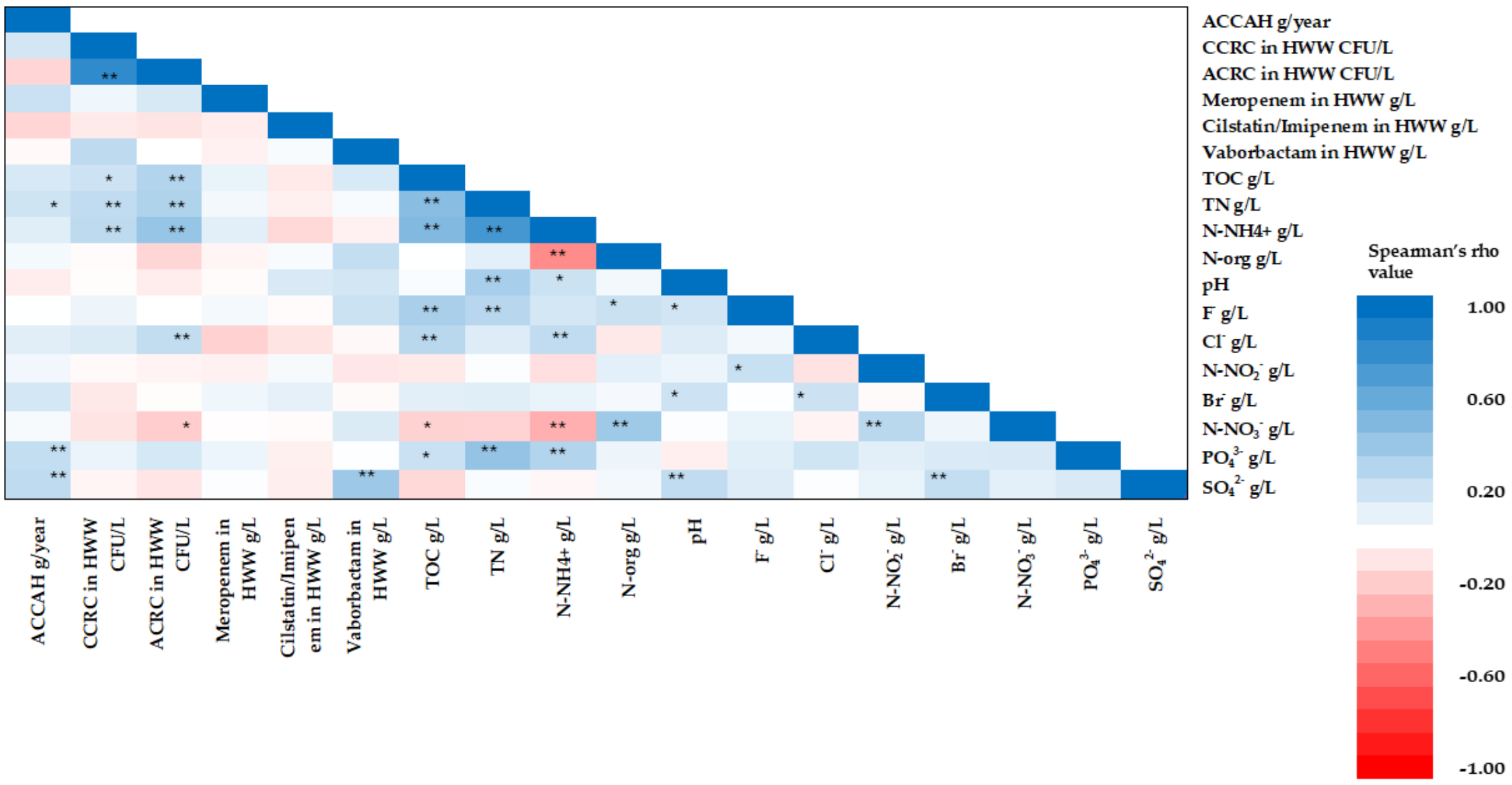

**Figure 7.** Heatmap presenting the values of Spearman's correlation coefficients. The highest values are marked in blue, and the lowest values are marked in red. One star denotes correlations that are significant at the 0.05 level (2-tailed), and two stars denote correlations that are significant at the 0.01 level (2-tailed). Level of significance: 0.05. Abbreviations: ACCAH – annual consumption of carbapenem antibiotics in hospitals; CCRC in HWW – cream-colored carbapenem-resistant colonies in hospital wastewater; ACRC in HWW – all carbapenem-resistant colonies in hospital wastewater. Source: own elaboration in IBM SPSS Statistics and MS Office.

In the group of the analyzed physicochemical parameters, positive correlations were noted between the concentrations of TOC, TN, and ammonium ions ($N\text{-}NH_4^+$) and between $N\text{-}NO_3^-$ and bromide ($Br^-$) ions. Total nitrogen content was bound by a positive correlation with $N\text{-}NH_4^+$, $N\text{-}NO_3^-$, and $Br^-$ ions, and by a weaker correlation with $N\text{-}NO_2^-$ions. The concentration of $N\text{-}NH_4^+$ ions was positively correlated with $N\text{-}NO_3^-$ and $Br^-$ ions, and negatively correlated with pH. The concentration of $Br^-$ ions was positively correlated with $N\text{-}NO_2^-$ ions. pH was negatively correlated with the concentration of $N\text{-}NO_3^-$ ions. These findings indicate that HWW contains various types of organic and inorganic pollutants and that different relationships exist between chemical parameters, for instance pH affects the concentrations of different forms of ammonia nitrogen. The observed correlations between organic matter decomposition and stages of the nitrogen cycle, between different forms of nitrogen and the pH of HWW, and between bromides and different forms of nitrogen and organic matter suggest that nutrient availability, environmental conditions, and the presence of specific ions can modulate the growth and metabolism of CRB.

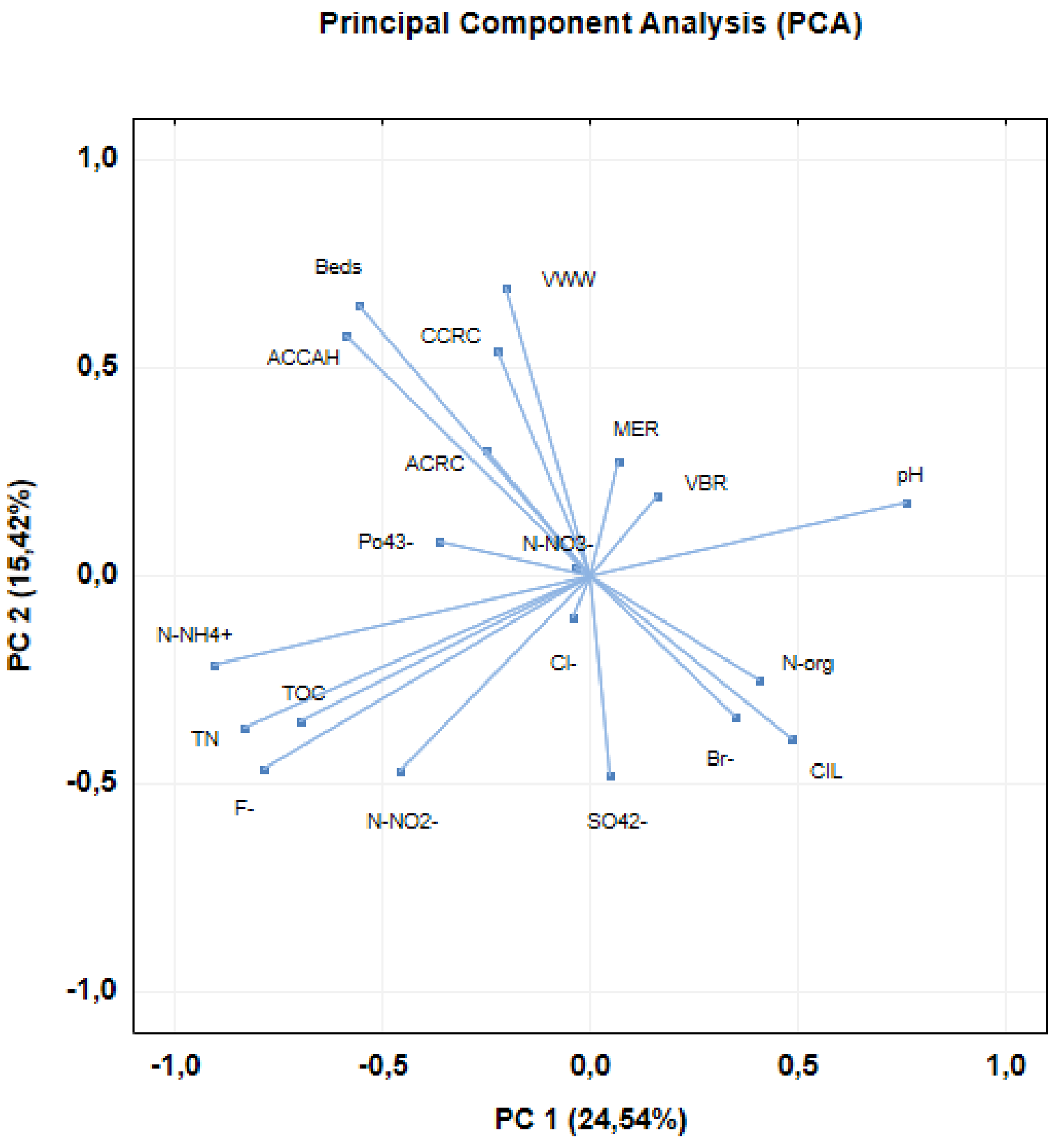

**Figure 8.** Principal component analysis of the key hospital data and the parameters analyzed in hospital wastewater samples. The following parameters were analyzed in HWW samples: ACRC – all carbapenem-resistant colonies, CCRC – cream-colored carbapenem-resistant colonies, MER – meropenem, VBR – vaborbactam, CIL / IMI – cilstatin/imipenem, concentration of ions ($Cl^-$, $Br^-$, $F^-$, $SO_4^{2-}$, N-$NO^{2-}$, N-$NO_3^-$, $PO_4^{3-}$, N-$NH_4^+$), TOC – total organic carbon, TN – total nitrogen, $N_{org}$ – organic nitrogen, and pH. Hospital data: ACCAH – annual consumption of carbapenem antibiotics in hospitals, Beds – bed capacity, VWW- daily volume of evacuated wastewater.

The PCA involved hospital data (bed capacity, volume of evacuated wastewater (VWW), and ACCAH) and the results of HWW analyses (physicochemical parameters and concentrations of carbapenem antibiotics). The results were used to determine seven principal components that explained more than 80% of total variance (Figure S5, Table S8). The first two principal components (PC1 and PC2) explained 39.96% of total variance (Figure 8). The first principal component (PC1, 24.54%) reflected the effects of variables such as TN, TOC, N-$NH_4^+$, VWW, and bed capacity, indicating that total CRB counts tend to increase in response to high concentrations of biogenic elements and in large hospitals. The second principal component (PC2, 15.42%) was correlated mainly with antibiotic concentrations in HWW (meropenem, cilastatin/imipenem) and microbial colonization levels, including CCRC counts. These relationships suggest that high antibiotic pressure and the presence of antibiotic residues in HWW contribute to the selection for CCRC, including CRPA and CRAB. Therefore, an increase in CCRC counts could point to the presence of highly pathogenic MDR strains in HWW. These results underscore the influence of both environmental factors and clinical practices on the composition and CR of the HWW microbiome.

## 4. Discussion

In this study, a total of 128 hospital wastewater samples were analyzed, representing 64 distinct healthcare facilities distributed across all 16 Polish voivodeships. The sampling campaign, conducted in both winter and summer seasons, allowed for a comprehensive mapping of carbapenem-resistant pathogens and antibiotic usage and chemical parameters in HWW at a national level. The findings of this study provide the first high-resolution, nationwide map of carbapenem resistance in Polish hospital effluents. Beyond the geographic scale, this study advances the conceptual framework of AMR surveillance by demonstrating that the HWW matrix acts not merely as a carrier, but as a selective incubator where specific physicochemical conditions (TOC, TN) and sub-inhibitory antibiotic and enzyme inhibitor concentrations (particularly vaborbactam) synergistically modulate the persistence of high-priority pathogens. By integrating data from 64 facilities across all administrative regions (voivodeships), we demonstrate that hospital wastewater is not merely a local concern but a significant national vector for the dissemination of high-priority pathogens (CRAB and CRPA) into the municipal sewage infrastructure. The present study demonstrated that the rise in antimicrobial resistance, in particular resistance to carbapenems that are often used as last-resort drugs, poses a severe and escalating threat to public health. Special attention was paid to critical and high priority pathogens such as CRAB and CRPA that present a substantial challenge in the clinical setting and play an important role in the dissemination of ARGs in the natural environment. The epidemiological and statistical data collated by the European Center for Disease Prevention and Control (https://www.ecdc.europa.eu/en), Centers for Disease Control and Prevention (Centers for Disease Control and Prevention (CDC), 2019), and the Pfizer Surveillance Atlas (https://atlas-surveillance.com) clearly indicate that CR is a serious global issue, particularly in bacteria of the genera *Acinetobacter* and *Pseudomonas* that cause drug-resistant infections and contribute to mortality. Talat et al. (2023) reported on the high prevalence of *A. baumannii*, *P. aeruginosa*, and other bacteria in HWW in India, and concluded that these microorganisms play a fundamental role in the spread of genes encoding resistance to various classes of antibiotics, which underscores the scale of the problem. Similar observations were made in a study of HWW discharged to the Mindu River in Manaus, Brazil, where MDR CRPA strains were identified in samples of HWW and river water at effluent discharge points, suggesting that these bacteria are able to survive in the aquatic environment (Magalhães et al., 2016). Aristizabal-Hoyos et al. (2023) also identified carbapenem-resistant Gram-negative bacteria in HWW in Latin America (Aristizabal-Hoyos et al., 2023). Analyses of HWW as a reservoir of antibiotic resistance in the natural environment are crucial because they enable scientists to monitor the circulation of ARGs outside the clinical setting. Zhang et al. (2020) identified carbapenemases such as NDM-1 in river water in China, which suggests that ARGs are transferred from healthcare facilities to natural water systems and pose a significant public health threat (Zhang et al., 2020). It should be emphasized that CR in the HWW microbiome affects not only long-existing, but also newly built hospitals. According to Boutin et al. (2024), CRB can be transmitted by person-to-person contact between patients and healthcare workers, and by acquisition from contaminated surfaces and HWW, regardless of the hospital's size or lifespan. The current study confirmed the presence of CRPA and CRAB in HWW discharged by Polish hospitals. Carbapenem-resistant *P. aeruginosa* strains were predominant, and they were detected in HWW samples from all Polish voivodeships. The highest CRPA counts were observed in HWW samples from large hospitals in highly urbanized regions. The prevalence of CRB was lower in smaller hospitals, which suggest that the microbial load in wastewater is closely linked to hospital size. The ERIC-PCR assay revealed considerable genetic diversity across CRPA strains, which could be indicative of clonal groups and unique variants. This observation suggests that the spread of drug resistance is a local and complex phenomenon. These results corroborate the findings of other researchers who also reported high levels of genetic variation across *P. aeruginosa* strains in hospitals and wastewater, which could result from adaptation to different ecological niches as well as horizontal transfer of ARGs (Porras-Agüera et al., 2020). In the analyzed CRPA strains, the average value of the clonal similarity index (0.18) and the presence of more than 30 main clusters point to considerable genetic variability, which hinders the identification of individual dominant clones, but also confirms that the dissemination of drug resistance is a multifaceted problem.

The proportion of CRAB strains in the total number of CCRC was relatively low - these strains were identified sporadically and did not form dominant groups. In contrast, Gu et al. (2004) found that CRAB was the dominant species in untreated HWW in eastern China (Gu et al., 2024). The low levels of genotypic similarity in CRAB (maximum genetic similarity – 50%, average clonal similarity – 0.09) could suggest that these strains had a different origin and were not actively transmitted between hospitals. This is an interesting observation, especially in the context of literature data which indicate that two CRAB clones – GC1 and GC2 – are globally dominant (Hamidian and Nigro, 2019). Crucially, our genomic analysis (ERIC-PCR) challenges the common narrative of global clone dominance (GC1/GC2) in hospital environments. The high genetic heterogeneity observed in isolates across Poland suggests that the carbapenem resistance ‘reservoir’ is maintained by a complex array of local, site-specific strains rather than a single epidemic lineage. This indicates that infection control strategies must be tailored to individual hospital ecologies rather than focusing solely on intercepting international clones. The differences in the dominant CRAB clones across geographic regions may be attributed to local selection pressures, antibiotic policies, differences in the epidemiology of healthcare-associated infections, and biofilm production. According to research, the genetic diversity of *A. baumannii* could be linked to their ability to adapt and acquire new ARGs, which plays a particularly important role in environments with high antibiotic loads, such as hospitals (Cabral et al., 2011). It should be noted that both CRPA and CRAB may play a pivotal role in the spread of antibiotic resistance, in particular CR, and that these pathogens can adapt to and persist in the clinical setting and the natural environment for a long time (Męcik et al., 2024). The geographic variations in CRB counts in HWW samples were linked to the specificity of the examined hospitals, and these pathogens were less prevalent in wastewater sampled from smaller hospitals with lower antibiotic consumption and lower daily volume of evacuated wastewater. The observed differences could be attributed to numerous factors, including epidemiological and demographic factors, awareness of hospital personnel, environmental conditions, availability of carbapenems to patients, and the standard of living in a given region or country (Gholipour et al., 2024; Stefaniak et al., 2024). In addition, last-resort antibiotics are more often prescribed in densely populated urban regions with better access to healthcare, which increases the risk of transmission of resistant bacterial strains. The significant positive correlation observed between annual carbapenems consumption in hospitals and the prevalence of CRAB provides empirical evidence that hospital prescribing patterns have a direct and measurable environmental footprint. This 'spillover effect' underscores that the clinical use of the last-resort antibiotics necessitates a corresponding environmental responsibility, specifically the implementation of advanced on-site pre-treatment technologies. This suggests that the dissemination of antibiotic resistance is influenced by both person-to-person contact and systemic factors (Bruinsma et al., 2003; Ternhag et al., 2014).

According to the latest research, hospitals are hotspots for the spread of antimicrobial resistance, and HWW is highly abundant in ARGs, pharmaceuticals, pathogenic and opportunistic bacteria that are transmitted to the environment (Silvester et al., 2025). The high variability of the HWW environment, conditioned by physicochemical factors (such as TOC, TN, and pH) and the presence of pharmaceuticals, can significantly affect the composition and dynamics of the wastewater microbiome as well as antibiotic resistance levels (Sharma et al., 2025). In the present study, CRPA strains were more frequently identified in HWW than CRAB strains, but CRAB strains were more genetically diverse in terms of the carried ARGs. The prevalence of the examined pathogens in HWW samples also differed across seasons. The number of CRAB strains harboring *sul*2, *bla*$_{CTX}$, and *bla*$_{OXA23/24/51\text{-like}}$ genes was higher in winter samples, which could be due to higher consumption of sulfonamides in this season. In CRPA strains, the abundance of ARGs, including *tet*A and *tet*M, was also higher in winter, which confirms seasonal variations in selection pressure.  In both cases (CRAB and CRPA), *bla*$_{NDM}$ was the most common MBL gene responsible for carbapenem resistance, but its frequency depended on the season in which the HWW samples were collected. The general frequency of *bla*$_{NDM}$ is consistent with the global literature (Mack et al., 2025a, 2025b). Seasonal differences in the prevalence of ARGs and CRGs were also found in HWW in India, which validates the observation that selection pressure fluctuates across seasons and that HWW is a key source of antibiotic resistance in the environment, especially in winter, when the consumption of antibiotics increases (Lamba et al., 2017). In addition, genetic variability can be associated with differences in temperature, antibiotic consumption, and the chemical composition of HWW across seasons, which can affect the survival and selection of resistant strains. These dependencies were confirmed by Shen et al. (2022) who studied the microbial diversity of municipal wastewater (Shen et al., 2022). The observed seasonal changes in resistance gene profiles highlights a critical temporal dimension in ARG dynamics. This periodicity suggests that environmental factors, such as temperature fluctuations in sewage systems, may catalyze horizontal gene transfer (HGT), potentially turning hospital drains into 'bioreactors' for resistance during warmer months. However, it should be noted that municipal wastewater is characterized by much lower concentrations of pharmaceuticals, in particular antibiotics, and different composition of microbial communities than HWW (Beattie et al., 2020; Souza and Féris, 2016).

Statistical analyses confirmed the presence of significant correlations between CRB counts and the physicochemical parameters of HWW, including TOC, TN, and vaborbactam concentration, which suggests that these parameters directly influence the abundance and survival of resistant microorganisms. The statistical correlations identified between these parameters and CRB prevalence reflect an underlying selective mechanism rather than mere coincidence. As highlighted by Rozman et al. (Rozman et al., 2020) hospital wastewater acts as a 'hot spot' where the constant influx of antibiotics, even at sub-inhibitory levels, creates a stable niche for the persistence of multidrug-resistant strains. This environment facilitates the maintenance of resistance determinants that might otherwise be lost due to fitness costs in less nutrient-rich settings. While the present study offers a nationwide map of CRB, its primary scientific advancement lies in the conceptualization of HWW as a dynamic biochemical incubator. Our PCA results indicate that organic enrichment (TOC/TN) and sub-inhibitory antibiotic and enzyme inhibitor concentrations (specifically meropenem and vaborbactam) act synergistically to maintain a high level of genomic plasticity. This is evidenced by the identification of over 30 distinct genetic clusters of *P. aeruginosa,* suggesting that HWW facilitates the emergence of local resistance variants rather than mere clonal dissemination. This finding aligns with the 'hotspot potential' framework proposed in recent international literature (Rozman et al., 2020; Secker et al., 2026) where the environmental matrix actively shapes the resistome before it reaches municipal treatment plants. To quantitatively contextualize these risks, the detected concentrations of meropenem and vaborbactam should be viewed in relation to the metabolic demands of the microbial community. The organic load (TOC/TN) acts as a synergistic factor; while antibiotics provide the selective pressure, the high nutrient availability allows resistant populations to thrive without the metabolic constraints typically associated with maintaining complex resistance mechanisms, such as carbapenemases. This mechanistic interplay explains the environmental persistence of CRPA and CRAB throughout both analyzed seasons, consistent with the 'pathogen reservoir' model (Secker et al., 2026), positioning HWW as a critical point-source for the dissemination of high-priority pathogens into urban infrastructure. In addition, the current study also demonstrated that HWW is a rich source of genes that encode resistance to numerous antibiotics, including carbapenems, MBLs, macrolides, sulfonamides, and tetracyclines, which indicates that HWW is a reservoir of genetic elements responsible for antimicrobial resistance. The fact that 50% of CRAB isolates and 25.45% of CRPA isolates carried multiple ARGs emphasizes the severity of the problem and suggests that comprehensive strategies are needed to control the spread of antimicrobial resistance. Recent research has shown that the variable environment of HWW can induce changes in dominant bacterial species and their genetic profile, including the presence of mobile genetic elements (MGEs) and integron classes responsible for the horizontal gene transfer (HGT) of ARGs (Keer et al., 2025; Silvester et al., 2025). The high abundance of *bla* genes, including CRGs such as OXA variants and MBLs encoded by *bla*$_{VIM}$ genes, is particularly alarming (Goudarzi et al., 2015; Poirel et al., 2011). According to Jafari-Sales et al. (2024) the *bla*$_{BIC}$ gene is one of the least clinically prevalent genes conferring resistance to carbapenems. Its detection in 16 CRPA strains originating from hospital wastewater in the present study may be particularly alarming from a One Health perspective. The presence of *bla*$_{VIM}$, *bla*$_{CTX}$, and *bla*$_{TEM}$ genes and variants of the *bla*$_{OXA}$ gene in CRAB and CRPA strains, as well as seasonal fluctuations in their prevalence, suggest that HWW is a dynamic environment that promotes the acquisition and dissemination of ARGs. It should be noted that MGEs play a key role in this dynamic environment by facilitating the rapid spread of ARGs between bacterial species. To strengthen the quantitative interpretation of environmental risk, the detected concentrations of meropenem and vaborbactam must be evaluated against established microbiological thresholds. The frequent detection of these compounds across 64 facilities, combined with the presence of multi-drug resistant CRPA and CRAB, suggests that HWW acts as a continuous source of selective pressure. According to the 'hot spot' model (Rozman et al., 2020), even nanogram-to-microgram per liter concentrations are sufficient to drive the selection of resistance genes within the complex biofilm of sewer systems. Quantitatively, the high prevalence of *bla*$_{NDM}$ and *bla*$_{VIM}$ genes identified in our seasonal HWW samples confirms that hospital effluents provide a substantial genetic load to the municipal infrastructure, transforming HWW into a measurable pathogen reservoir (Secker et al., 2026) that necessitates specific effluent discharge limits to mitigate the risk of AMR dissemination. In conclusion, our results demonstrate that current wastewater management protocols are insufficient to contain the environmental flux of carbapenem resistance. This study serves as a critical call to action for policymakers to treat hospital wastewater as a high-risk industrial byproduct, requiring mandatory disinfection before it enters the public sewer system to protect both environmental integrity and public health.

## 5. Conclusions

The current study confirmed the presence of CRPA and CRAB strains in HWW in Poland and demonstrated that CRPA strains were predominant in the examined HWW samples. Carbapenem-resistant *P. aeruginosa* were isolated more frequently than CRAB and were characterized by genetic heterogeneity, which suggests the presence of many independent clonal lines. In turn, CRAB strains were identified less frequently, but they were also characterized by genotypic variability, which could indicate that these pathogens spread locally from many point sources. Carbapenem-resistant bacteria were far more prevalent in wastewater from large hospitals and highly urbanized regions. The isolated bacterial strains harbored numerous ARGs, including CRGs such as ESBLs ($bla_{OXA}$, $bla_{CTX}$, $bla_{TEM}$, $bla_{SHV}$), MBLs ($bla_{VIM}$, $bla_{NDM}$, $bla_{IMP}$), and other *bla* genes ($bla_{OXA23\text{-}like}$, $bla_{OXA24\text{-}like}$, $bla_{OXA51\text{-}like}$, $bla_{OXA58\text{-}like}$, $bla_{BIC}$, $bla_{SPM}$, $bla_{AIM}$, $bla_{GIM}$), as well as genes encoding resistance to other antibiotics (*sul*1, *sul*2, *tet*M, *erm*F, *qep*A). The analyzed HWW samples were also colonized by strains with CRGs and other multiple ARGs in both *A. baumannii* and *P. aeruginosa*. The abundance of CRB was significantly correlated with selected physicochemical parameters of HWW, including TOC, TN, and vaborbactam concentration, which indicates that environmental conditions affect the survival and selection of drug-resistant strains. The results of this study underscore the urgent need for effective methods of monitoring chemical and biological contamination of HWW as well as effective wastewater treatment techniques. The present findings indicate that HWW should be regularly monitored and that microbial pollution control strategies consistent with the One Health approach should be implemented to integrate human, animal, and environmental health. The prescription of antibiotics in healthcare facilities should also be monitored and rationalized as a key element of the strategy to combat the spread of antimicrobial resistance.

**Future prospects and limitations of this study**

The main aim of the study was to determine the scale of the problem of carbapenem-resistant PA and AB prevalence in Poland, considering the level of carbapenem antibiotic administration in different regions of Poland. The low recovery of PA and AB from the CHROMagar mSuperCARBA medium may have been due to the fact that only non-Enterobacterales carbapenem-resistant bacteria were selected to analyses. It was considered to use MALDI-TOF to identify the strains, yet preliminary studies and previously published results of own studies found this method of bacteria identification as ineffective in environmental samples (Osińska et al., 2023). Further analysis of diversity/similarity of individual strains will be the subject of a separate study currently in preparation, based on the structure of the entire metagenome (using the WGS method).

**Author contribution statement**
**Męcik Magdalena:** Conceptualization, Validation, Data Curation, Formal analysis, Data curation Writing - Original draft, Sampling collection, Visualization, **Kornelia Stefaniak:** Formal analysis, Sampling collection, Review, **Harnisz Monika:** Conceptualization, Review, Sampling collection, **Sylwia Bajkacz**: Review, Sampling collection, Formal Analysis, Data curation; **Ewa Felis:** Review, Sampling collection, Formal Analysis, Data curation; **Joanna Wilk:** Formal Chemical Analysis, Data curation **Karolina Dudek:** Formal analysis, data curation **Korzeniewska Ewa:** Conceptualization, Methodology, Validation, Resources, Sampling collection, Supervision, Project administration, Funding acquisition.

**Declaration of Competing Interest**
The authors declare that they have no known competing financial interests or personal relationships that could have influenced the work reported in this paper.

**Acknowledgements**
This work was supported by the Polish National Science Center [Grant No. 2022/45/B/NZ7/00793]. For the purpose of Open Access, the author has applied a CC-BY-SA 4.0 public copyright license to any Author Accepted Manuscript (AAM) version arising from this submission.

**Supplementary data**
Supplementary data for this article can be found at:

**References**
Agarwal, S., Darbar, S., Saha, S., 2022. Challenges in management of domestic wastewater for sustainable development 6, 531–552. https://doi.org/10.1016/B978-0-323-91838-1.00019-1

Al-Ahmadi, J., Roodsari, Z., 2016. Fast and Specific Detection of Pseudomonas Aeruginosa From Other Pseudomonas Species By Pcr. Ann. Burns Fire Disasters XXIX, 5–8.

Amudhan, S.M., Sekar, U., Arunagiri, K., Sekar, B., 2011. OXA betaβ-lactamase-mediated carbapenem resistance in Acinetobacter baumannii. Indian J. Med. Microbiol. 29, 269–274. https://doi.org/10.4103/0255-0857.83911

Aristizabal-Hoyos, A.M., Rodríguez, E.A., Torres-Palma, R.A., Jiménez, J.N., 2023. Concern levels of beta-lactamase-producing Gram-negative bacilli in hospital wastewater: hotspot of antimicrobial resistance in Latin-America. Diagn. Microbiol. Infect. Dis. 105. https://doi.org/10.1016/j.diagmicrobio.2022.115819

Beattie, R.E., Skwor, T., Hristova, K.R., 2020. Survivor microbial populations in post-chlorinated wastewater are strongly associated with untreated hospital sewage and include ceftazidime and meropenem resistant populations. Sci. Total Environ. 740, 140186. https://doi.org/10.1016/j.scitotenv.2020.140186

Boutin, S., Scherrer, M., Späth, I., Kocer, K., Heeg, K., Nurjadi, D., 2024. Cross-contamination of carbapenem-resistant Gram-negative bacteria between patients and the hospital environment in the first year of a newly built surgical ward. J. Hosp. Infect. 144, 118–127. https://doi.org/10.1016/j.jhin.2023.11.016

Bruinsma, N., Hutchinson, J.M., van den Bogaard, A.E., Giamarellou, H., Degener, J., Stobberingh, E.E., 2003. Influence of population density on antibiotic resistance. J. Antimicrob. Chemother. 51, 385–390. https://doi.org/10.1093/jac/dkg072

Cabral, M.P., Soares, N.C., Aranda, J., Parreira, J.R., Rumbo, C., Poza, M., Valle, J., Calamia, V., Lasa, Í., Bou, G., 2011. Proteomic and functional analyses reveal a unique lifestyle for acinetobacter baumannii biofilms and a key role for histidine metabolism. J. Proteome Res. 10, 3399–3417. https://doi.org/10.1021/pr101299j

Carraro, E., Bonetta, Silvia, Bonetta, Sara, 2018. Hospital wastewater: Existing regulations and current trends in management. Handb. Environ. Chem. 60, 1–16. https://doi.org/10.1007/698_2017_10

Catalano, A., Iacopetta, D., Ceramella, J., Scumaci, D., Giuzio, F., Saturnino, C., Aquaro, S., Rosano, C., Sinicropi,

M.S., 2022. Multidrug Resistance (MDR): A Widespread Phenomenon in Pharmacological Therapies. Molecules 27, 1–18. https://doi.org/10.3390/molecules27030616

Centers for Disease Control and Prevention (CDC), 2019. Antibiotic resistance threats in the United States, 2019. Atlanta, Georgia. https://doi.org/10.15620/cdc:82532

Chen, L., Zhou, H., Yu, B., Huang, Z.W., 2014. Comparison study on hospital wastewater disinfection technology. Adv. Mater. Res. 884–885, 41–45. https://doi.org/10.4028/www.scientific.net/AMR.884-885.41

Chen, T.L., Siu, L.K., Wu, R.C.C., Shaio, M.F., Huang, L.Y., Fung, C.P., Lee, C.M., Cho, W.L., 2007. Comparison of one-tube multiplex PCR, automated ribotyping and intergenic spacer (ITS) sequencing for rapid identification of Acinetobacter baumannii. Clin. Microbiol. Infect. 13, 801–806. https://doi.org/10.1111/j.1469-0691.2007.01744.x

Dashti, A.A., Jadaon, M.M., Abdulsamad, A.M., Dashti, H.M., 2009. Heat treatment of bacteria: A simple method of DNA extraction for molecular techniques. Kuwait Med. J. 41, 117–122.

De Vos, D., Lim, A., Pirnay, J.P., Struelens, M., Vandenvelde, C., Duinslaeger, L., Vanderkelen, A., Cornelis, P., 1997. Direct detection and identification of Pseudomonas aeruginosa in clinical samples such as skin biopsy specimens and expectorations by multiplex PCR based on two outer membrane lipoprotein genes, oprI and oprL. J. Clin. Microbiol. 35, 1295–1299. https://doi.org/10.1128/jcm.35.6.1295-1299.1997

Eitel, Z., Sóki, J., Urbán, E., Nagy, E., 2013. The prevalence of antibiotic resistance genes in Bacteroides fragilis group strains isolated in different European countries. Anaerobe 21, 43–49. https://doi.org/10.1016/j.anaerobe.2013.03.001

Gholipour, S., Nikaeen, M., Mohammadi, F., Rabbani, D., 2024. Antibiotic resistance pattern of waterborne causative agents of healthcare-associated infections: A call for biofilm control in hospital water systems. J. Infect. Public Health 17, 102469. https://doi.org/10.1016/j.jiph.2024.102469

Goldstein, C., Lee, M.D., Sanchez, S., Hudson, C., Phillips, B., Register, B., Grady, M., Liebert, C., Summers, A.O., White, D.G., Maurer, J.J., 2001. Incidence of class 1 and 2 integrases in clinical and commensal bacteria from livestock, companion animals, and exotics. Antimicrob. Agents Chemother. 45, 723–726. https://doi.org/10.1128/AAC.45.3.723-726.2001

Goudarzi, H., Mirsamadi, E.S., Ghalavand, Z., Vala, M.H., Mirjalali, H., Hashemi, A., Ghasemi, E., 2015. Molecular detection of metallo-beta-lactamase genes in clinical isolates of Acinetobacter baumannii. J. Pure Appl. Microbiol. 9, 145–151.

Gu, D., Wu, Y., Chen, K., Zhang, Y., Ju, X., Yan, Z., Xie, M., Chan, E.W.C., Chen, S., Ruan, Z., Zhang, R., Zhang, J., 2024. Recovery and genetic characterization of clinically-relevant ST2 carbapenem-resistant Acinetobacter baumannii isolates from untreated hospital sewage in Zhejiang Province, China. Sci. Total Environ. 916, 170058. https://doi.org/10.1016/j.scitotenv.2024.170058

GUS Główny Urząd Statystyczny, 2024. Powierzchnia i ludność w przekroju terytorialnym w 2024 roku. Warsaw.

GUS Główny Urząd Statystyczny, 2023a. Statistical Yearbook of the Republic of Poland 2023. Zakład Wydawnictw Statystycznych, Warsaw.

GUS Główny Urząd Statystyczny, 2023b. Rocznik Statystyczny Województw. Zakład Wydawnictw Statystycznych, Warsaw.

GUS Główny Urząd Statystyczny, 2022. Sytuacja społeczno-gospodarcza województw Nr 1 / 2022. Zakład Wydawnictw Statystycznych, Warsaw.

Hamidian, M., Nigro, S.J., 2019. Emergence, molecular mechanisms and global spread of carbapenem-resistant acinetobacter baumannii. Microb. Genomics 5. https://doi.org/10.1099/mgen.0.000306

Heras, J., Domínguez, C., Mata, E., Pascual, V., Lozano, C., Torres, C., Zarazaga, M., 2015. GelJ - a tool for analyzing DNA fingerprint gel images. BMC Bioinformatics 16, 1–8. https://doi.org/10.1186/s12859-015-0703-0

Hubeny, J., Korzeniewska, E., Buta-Hubeny, M., Zieliński, W., Rolbiecki, D., Harnisz, M., 2022. Characterization

of carbapenem resistance in environmental samples and Acinetobacter spp. isolates from wastewater and river water in Poland. Sci. Total Environ. 822. https://doi.org/10.1016/j.scitotenv.2022.153437

Jafari-sales, A., Mehdizadeh, F., Fallah, G., Pashazadeh, M., 2024. Examining the frequency of carbapenemase genes blaKPC, blaIMP, blaOXA-48, blaSPM, blaNDM, blaVIM, blaGES, blaBIC, blaAIM, blaGIM, blaSIM, and blaDIM in Pseudomonas aeruginosa strains isolated from patients hospitalized in Northwest Iran hospitals. J. Exp. Clin. Med. 41, 466–473. https://doi.org/10.52142/omujecm.41.3.3

Jean, S.S., Harnod, D., Hsueh, P.R., 2022. Global Threat of Carbapenem-Resistant Gram-Negative Bacteria. Front. Cell. Infect. Microbiol. 12, 1–19. https://doi.org/10.3389/fcimb.2022.823684

Jiang, Y., Ding, Y., Wei, Y., Jian, C., Liu, J., Zeng, Z., 2022. Carbapenem-resistant Acinetobacter baumannii: A challenge in the intensive care unit. Front. Microbiol. 13, 1–15. https://doi.org/10.3389/fmicb.2022.1045206

Keer, A., Oza, Y., Mongad, D., Ramakrishnan, D., Dhotre, D., Ahmed, A., Zumla, A., Shouche, Y., Sharma, A., 2025. Assessment of seasonal variations in antibiotic resistance genes and microbial communities in sewage treatment plants for public health monitoring. Environ. Pollut. 375, 126367. https://doi.org/10.1016/j.envpol.2025.126367

Khurana, P., Pulicharla, R., Brar, S.K., 2024. Occurrence of Imipenem in natural water: Effect of dissolved organic matter and metals. Sci. Total Environ. 957, 177846. https://doi.org/10.1016/j.scitotenv.2024.177846

Kim, J., Jeon, S., Rhie, H., Lee, B., Park, M., Lee, H., Lee, J., Kim, S., 2009. Rapid Detection of Extended Spectrum β-Lactamase (ESBL) for Enterobacteriaceae by use of a Multiplex PCR-based Method. Infect. Chemother. 41, 181. https://doi.org/10.3947/ic.2009.41.3.181

Kumari, A., Maurya, N.S., Tiwari, B., 2020. Hospital wastewater treatment scenario around the globe, in: Current Developments in Biotechnology and Bioengineering. Elsevier, Quebec City, pp. 549–570. https://doi.org/10.1016/B978-0-12-819722-6.00015-8

Lamba, M., Graham, D.W., Ahammad, S.Z., 2017. Hospital Wastewater Releases of Carbapenem-Resistance Pathogens and Genes in Urban India. Environ. Sci. Technol. 51, 13906–13912. https://doi.org/10.1021/acs.est.7b03380

Li, J., Wang, T., Shao, B., Shen, J., Wang, S., Wu, Y., 2012. Plasmid-mediated quinolone resistance genes and antibiotic residues in wastewater and soil adjacent to swine feedlots: Potential transfer to agricultural lands. Environ. Health Perspect. 120, 1144–1149. https://doi.org/10.1289/ehp.1104776

Ma, Y., Wu, N., Zhang, T., Li, Y., Cao, L., Zhang, P., Zhang, Z., Zhu, T., Zhang, C., 2024. The microbiome, resistome, and their co-evolution in sewage at a hospital for infectious diseases in Shanghai, China. Microbiol. Spectr. 12. https://doi.org/10.1128/spectrum.03900-23

Mack, A.R., Hujer, A.M., Mojica, M.F., Taracila, M.A., Feldgarden, M., Haft, D.H., 2025a. β-Lactamase diversity in Acinetobacter baumannii. Antimicrob. Chemother. 69. https://doi.org/https://journals.asm.org/doi/10.1128/aac.00785-24

Mack, A.R., Hujer, A.M., Mojica, M.F., Taracila, M.A., Feldgarden, M., Haft, D.H., 2025b. β-Lactamase diversity in Pseudomonas aeruginosa. Antimicrob. Chemother. https://doi.org/https://doi.org/10.1128/aac.00784-24

Magalhães, M.J.T.L., Pontes, G., Serra, P.T., Balieiro, A., Castro, D., Pieri, F.A., Crainey, J.L., Nogueira, P.A., Orlandi, P.P., 2016. Multidrug resistant Pseudomonas aeruginosa survey in a stream receiving effluents from ineffective wastewater hospital plants. BMC Microbiol. 16, 1–8. https://doi.org/10.1186/s12866-016-0798-0

Męcik, M., Stefaniak, K., Harnisz, M., Korzeniewska, E., 2024. Hospital and municipal wastewater as a source of carbapenem-resistant Acinetobacter baumannii and Pseudomonas aeruginosa in the environment: a review. Environ. Sci. Pollut. Res. 31, 48813–48838. https://doi.org/10.1007/s11356-024-34436-x

Mourabiti, F., Jouga, F., Sakoui, S., El Hosayny, O., Zouheir, Y., Soukri, A., El Khalfi, B., 2025. Mechanisms, therapeutic strategies, and emerging therapeutic alternatives for carbapenem resistance in Gram-negative bacteria. Arch. Microbiol. 207, 58. https://doi.org/10.1007/s00203-025-04252-z

Ng, L.K., Martin, I., Alfa, M., Mulvey, M., 2001. Multiplex PCR for the detection of tetracycline resistant genes.

Mol. Cell. Probes 15, 209–215. https://doi.org/10.1006/mcpr.2001.0363

Osińska, A., Harnisz, M., Korzeniewska, E., 2016. Prevalence of plasmid-mediated multidrug resistance determinants in fluoroquinolone-resistant bacteria isolated from sewage and surface water. Environ. Sci. Pollut. Res. 23, 10818–10831. https://doi.org/10.1007/s11356-016-6221-4

Osińska, A., Korzeniewska, E., Harnisz, M., Niestępski, S., 2017. The prevalence and characterization of antibiotic-resistant and virulent Escherichia coli strains in the municipal wastewater system and their environmental fate. Sci. Total Environ. 577, 367–375. https://doi.org/10.1016/j.scitotenv.2016.10.203

Osińska, A., Korzeniewska, E., Korzeniowska, A., Anna, K., Monika, W., 2023. The challenges in the identification of Escherichia coli from environmental samples and their genetic characterization. Environ. Sci. Pollut. Res. 11572–11583. https://doi.org/10.1007/s11356-022-22870-8

Pei, R., Kim, S.C., Carlson, K.H., Pruden, A., 2006. Effect of River Landscape on the sediment concentrations of antibiotics and corresponding antibiotic resistance genes (ARG). Water Res. 40, 2427–2435. https://doi.org/10.1016/j.watres.2006.04.017

Perry, M.R., Lepper, H.C., McNally, L., Wee, B.A., Munk, P., Warr, A., Moore, B., Kalima, P., Philip, C., de Roda Husman, A.M., Aarestrup, F.M., Woolhouse, M.E.J., van Bunnik, B.A.D., 2021. Secrets of the Hospital Underbelly: Patterns of Abundance of Antimicrobial Resistance Genes in Hospital Wastewater Vary by Specific Antimicrobial and Bacterial Family. Front. Microbiol. 12. https://doi.org/10.3389/fmicb.2021.703560

Poirel, L., Walsh, T.R., Cuvillier, V., Nordmann, P., 2011. Multiplex PCR for detection of acquired carbapenemase genes. Diagn. Microbiol. Infect. Dis. 70, 119–123. https://doi.org/10.1016/j.diagmicrobio.2010.12.002

Porras-Agüera, J.A., Moreno-García, J., González-Jiménez, M.D.C., Mauricio, J.C., Moreno, J., García-Martínez, T., 2020. Autophagic proteome in two saccharomyces cerevisiae strains during second fermentation for sparkling wine elaboration. Microorganisms 8. https://doi.org/10.3390/microorganisms8040523

Ramírez-Coronel, A.A., Mohammadi, M.J., Majdi, H.S., Zabibah, R.S., Taherian, M., Prasetio, D.B., Gabr, G.A., Asban, P., Kiani, A., Sarkohaki, S., 2024. Hospital wastewater treatment methods and its impact on human health and environments. Rev. Environ. Health 39, 423–434. https://doi.org/10.1515/reveh-2022-0216

Rolbiecki, D., Harnisz, M., Korzeniewska, E., Buta, M., Hubeny, J., Zieliński, W., 2021. Detection of carbapenemase-producing, hypervirulent Klebsiella spp. in wastewater and their potential transmission to river water and WWTP employees. Int. J. Hyg. Environ. Health 237. https://doi.org/10.1016/j.ijheh.2021.113831

Rozman, U., Duh, D., Cimerman, M., Turk, S.Š., 2020. Hospital wastewater effluent: Hot spot for antibiotic resistant bacteria. J. Water Sanit. Hyg. Dev. 10, 171–178. https://doi.org/10.2166/washdev.2020.086

Secker, B., Nayak, A., Husain, A.A., Arora, S., Nag, A., Shrivastava, S.K., Singer, A.C., Gomes, R.L., Acheampong, E., Chidambaram, S.B., Bhatnagar, T., Vetrivel, U., Kashyap, R.S., Atterbury, R.J., Blanchard, A.M., Monaghan, T.M., 2026. Metagenomic insights into the urban–rural variation of antimicrobial resistance and pathogen reservoirs in untreated wastewater from central India. Front. Microbiol. 16. https://doi.org/10.3389/fmicb.2025.1722229

Sharma, C., Gupta, S., Kumar, Vijay, Kumar, Vivek, 2025. Hospital-associated effluents: the masked environmental threat that needs urgent attention and action. Discov. Appl. Sci. 7. https://doi.org/10.1007/s42452-024-06456-2

Shen, W., Chen, Y., Wang, N., Wan, P., Peng, Z., Zhao, H., Wang, W., Xiong, L., Zhang, S., Liu, R., 2022. Seasonal variability of the correlation network of antibiotics, antibiotic resistance determinants, and bacteria in a wastewater treatment plant and receiving water. J. Environ. Manage. 317, 115362. https://doi.org/10.1016/j.jenvman.2022.115362

Sib, E., Voigt, A.M., Wilbring, G., Schreiber, C., Faerber, H.A., Skutlarek, D., Parcina, M., Mahn, R., Wolf, D., Brossart, P., Geiser, F., Engelhart, S., Exner, M., Bierbaum, G., Schmithausen, R.M., 2019. Antibiotic resistant bacteria and resistance genes in biofilms in clinical wastewater networks. Int. J. Hyg. Environ. Health 222, 655–662. https://doi.org/10.1016/j.ijheh.2019.03.006

Silvester, R., Perry, W.B., Webster, G., Rushton, L., Baldwin, A., Pass, D.A., Healey, N., Farkas, K., Craine, N., Cross, G., Kille, P., Weightman, A.J., Jones, D.L., 2025. Metagenomics unveils the role of hospitals and wastewater treatment plants on the environmental burden of antibiotic resistance genes and opportunistic pathogens. Sci. Total Environ. 961, 178403. https://doi.org/10.1016/j.scitotenv.2025.178403

Souza, F.S., Féris, L.A., 2016. Hospital and Municipal Wastewater: Identification of Relevant Pharmaceutical Compounds. Water Environ. Res. 88, 871–877. https://doi.org/10.2175/106143016x14609975747603

Stefaniak, K., Harnisz, M., Magdalena, M., Korzeniewska, E., 2025. ARB inactivation , ARGs and antibiotics degradation in hospital wastewater 495. https://doi.org/10.1016/j.jhazmat.2025.138833

Stefaniak, K., Kiedrzyński, M., Korzeniewska, E., Kiedrzyńska, E., Harnisz, M., 2024. Preliminary insights on carbapenem resistance in Enterobacteriaceae in high-income and low-/middle-income countries. Sci. Total Environ. 957. https://doi.org/10.1016/j.scitotenv.2024.177593

Talat, A., Blake, K.S., Dantas, G., Khan, A.U., 2023. Metagenomic Insight into Microbiome and Antibiotic Resistance Genes of High Clinical Concern in Urban and Rural Hospital Wastewater of Northern India Origin: a Major Reservoir of Antimicrobial Resistance. Microbiol. Spectr. 11. https://doi.org/10.1128/spectrum.04102-22

Ternhag, A., Grünewald, M., Nauclér, P., Tegmark Wisell, K., 2014. Antibiotic consumption in relation to socio-demographic factors, co-morbidity, and accessibility of primary health care. Scand. J. Infect. Dis. 46, 888–896. https://doi.org/10.3109/00365548.2014.954264

Turczak, A., 2017. Analiza przyczynowa różnic w poziomie PKB per capita między województwami w Polsce. Rocz. Ekon. i Zarządzania 9(45), 75–90. https://doi.org/10.18290/reiz.2017.9.4-4

Versalovic, J., Koeuth, T., Lupski, R., 1991. Distribution of repetitive DNA sequences in eubacteria and application to finerpriting of bacterial enomes. Nucleic Acids Res. 19, 6823–6831. https://doi.org/10.1093/nar/19.24.6823

Werkneh, A.A., Islam, M.A., 2023. Post-treatment disinfection technologies for sustainable removal of antibiotic residues and antimicrobial resistance bacteria from hospital wastewater. Heliyon 9, e15360. https://doi.org/10.1016/j.heliyon.2023.e15360

World Health Organization (WHO), 2024. WHO Bacterial Priority Pathogens List, 2024: bacterial pathogens of public health importance to guide research, development and strategies to prevent and control antimicrobial resistance, Bacterial pathogens of public health importance to guide research, development and strategies to prevent and control antimicrobial resistance.

Zhang, L., Ma, X., Luo, L., Hu, N., Duan, J., Tang, Z., Zhong, R., Li, Y., 2020. The prevalence and characterization of extended-spectrum β-lactamase-and carbapenemase-producing bacteria from hospital sewage, treated effluents and receiving rivers. Int. J. Environ. Res. Public Health 17. https://doi.org/10.3390/ijerph17041183

Zou, L., Meng, F., Hu, L., Huang, Q., Liu, M., Yin, T., 2019. A novel reversed-phase high-performance liquid chromatographic assay for the simultaneous determination of imipenem and meropenem in human plasma and its application in TDM. J. Pharm. Biomed. Anal. 169, 142–150. https://doi.org/10.1016/j.jpba.2019.01.039

# Supplementary Materials

# Tracking Carbapenem-Resistant Pathogens in Hospital Wastewater: the focus on *Acinetobacter baumannii* and *Pseudomonas aeruginosa*

**Magdalena Męcik [1], Kornelia Stefaniak [1], Monika Harnisz [1], Ewa Felis [2,3], Sylwia Bajkacz [3,4], Joanna Wilk[4], Karolina Dudek[1] and Ewa Korzeniewska [1]***

[1] Department of Water Protection Engineering and Environmental Microbiology, Faculty of Geoengineering, University of Warmia and Mazury in Olsztyn, Prawocheńskiego 1, 10-720 Olsztyn, Poland
[2] Silesian University of Technology, Faculty of Power and Environmental Engineering, Environmental Biotechnology Department, Akademicka 2 Str., Gliwice, 44-100, Poland
[3] Silesian University of Technology, Biotechnology Centre, B. Krzywoustego 8 Str., Gliwice, 44-100, Poland
[4] Silesian University of Technology, Faculty of Chemistry, Department of Inorganic Chemistry, Analytical Chemistry and Electrochemistry, B. Krzywoustego 6 Str., Gliwice, 44-100, Poland

* Corresponding author: ewa.korzeniewska@uwm.edu.pl

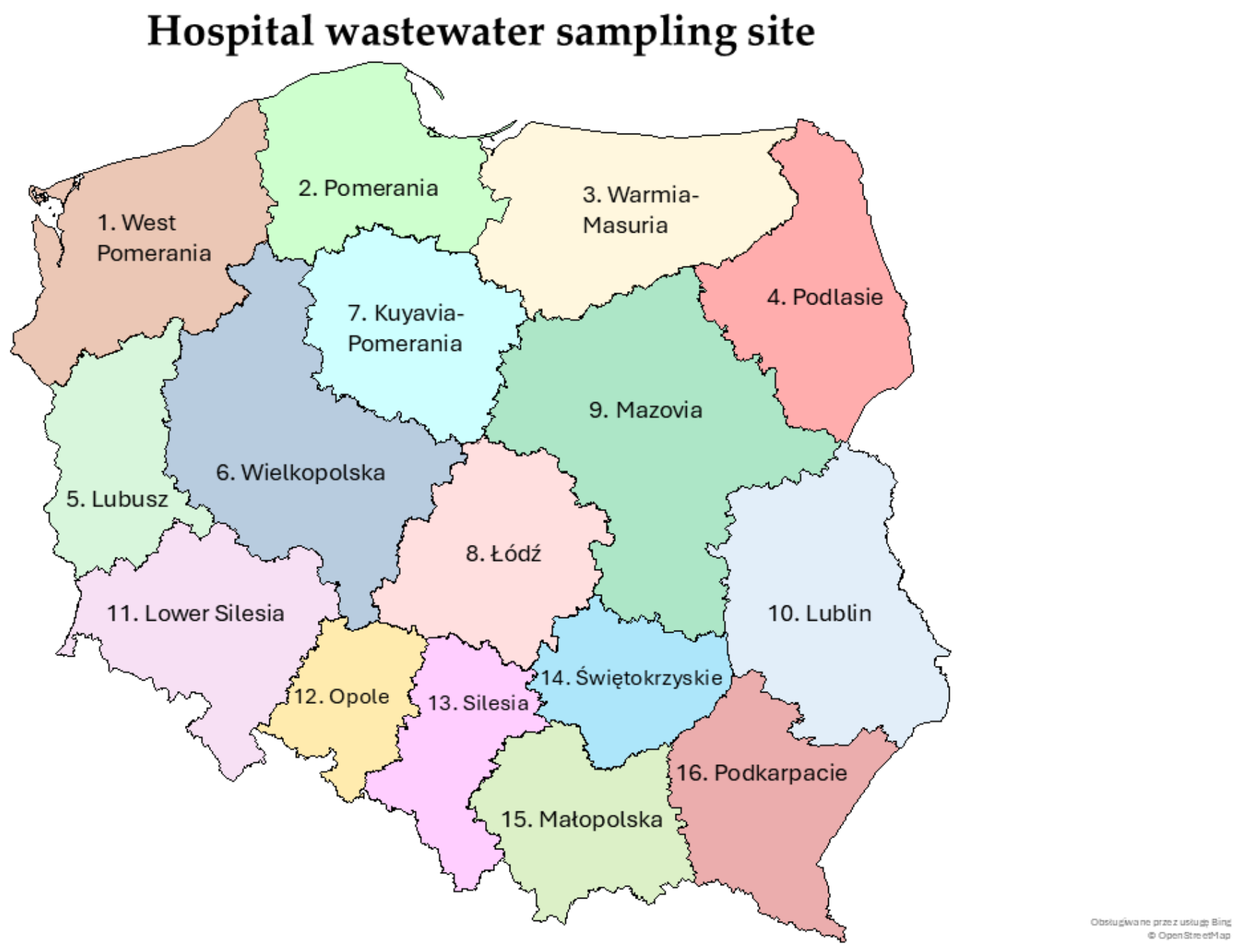

**Figure S1.** Voivodeships in Poland. Created with MS Office.

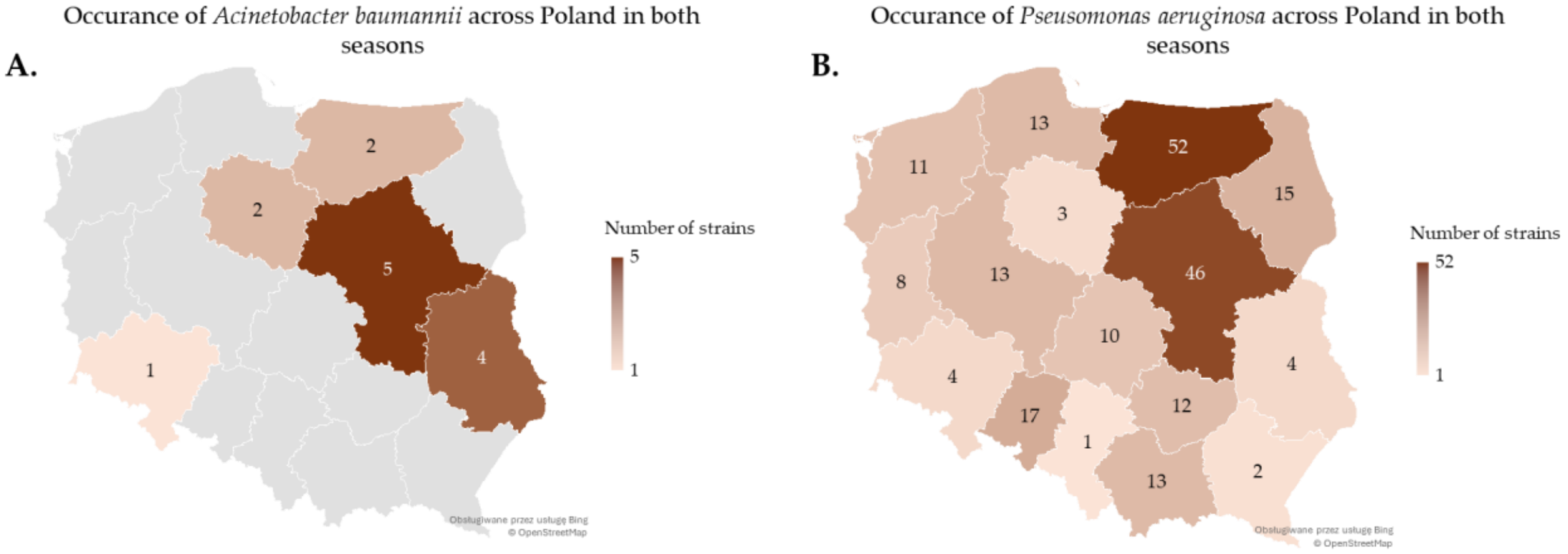

**Figure S2.** Heatmap of carbapenem-resistant *Acinetobacter baumannii* (A.) and *Pseudomonas aeruginosa* (B.) across Poland based on the number of collected strains in both sampling seasons. Voivodeships with the lowest percentage of resistant strains are marked in light brown; voivodeships with the highest percentage of resistant strains are marked in dark brown. Color gray indicates no collected strains of CRAB or CRPA in the area. Created with MS Office.

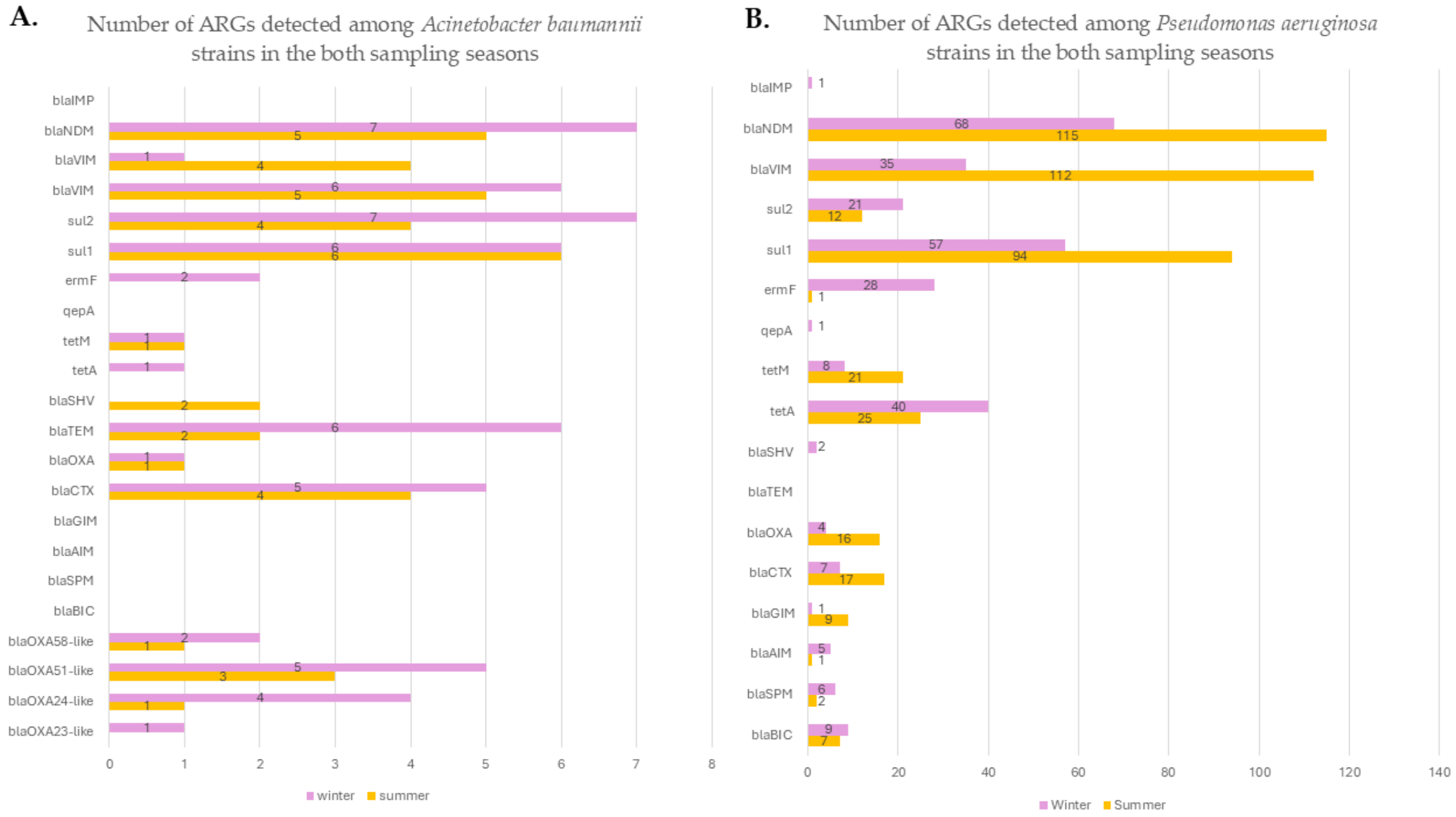

**Figure S3.**: Occurrence of all tested CRGs and other ARGs during each sampling season in CRAB (A.) and CRPA (B.).

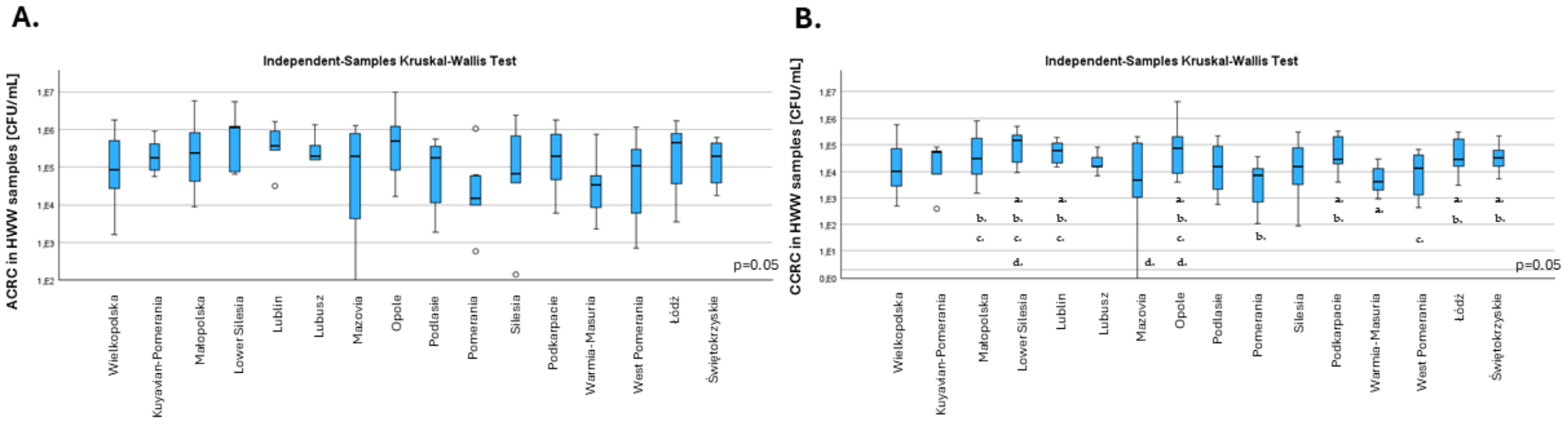

**Figure S4.** The difference between the frequency of occurrence of all carbapenem-resistant colonies (ACRC) (A.) (CFU/ml) and cream-colored carbapenem-resistant colonies (CCRC) (B.) collected from HWW samples by province of origin. Individual lowercase letters indicate statistically significant differences in the prevalence of CCRC in HWW samples between the provinces of Warmian-Masurian (a), Pomeranian (b), West Pomeranian (c) and Masovian (d) and the provinces marked with these letters in the chart, as well as the frequency of CCRC in HWW samples from other provinces. Created using IBM SPSS Statistics software.

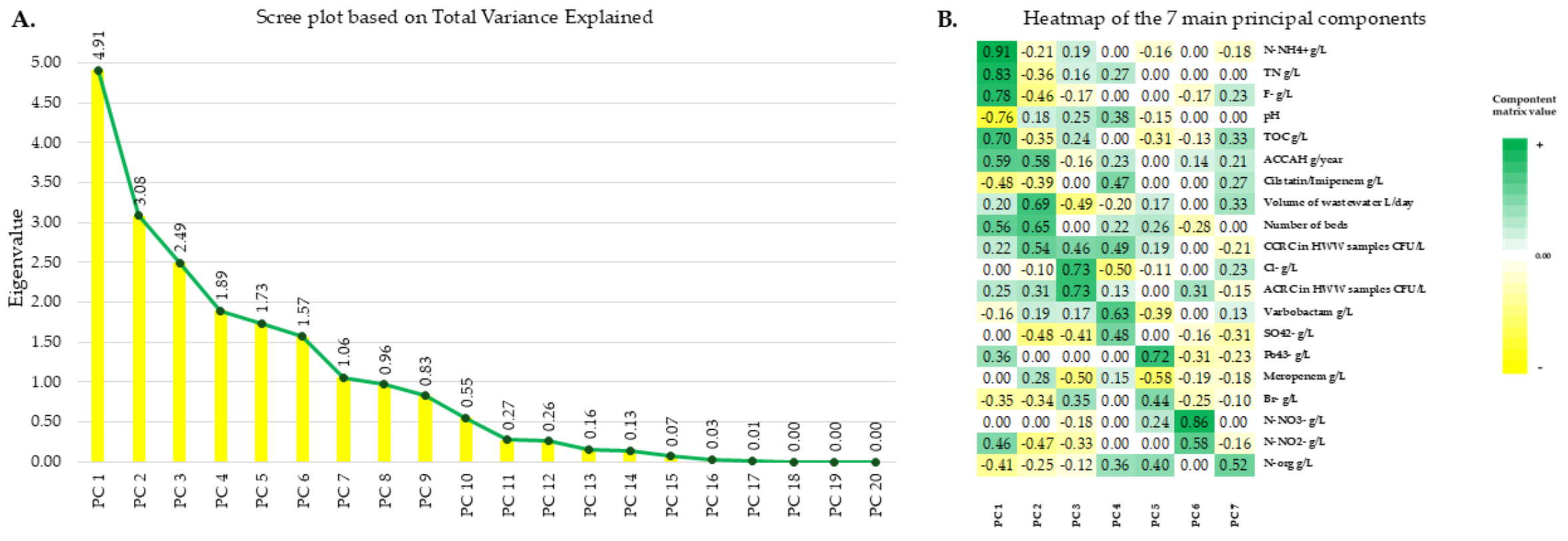

**Figure S5.** Principal Component Analysis. Scree plot (A.) of all found components and heatmap of component matrix value (B.) of 7 main principal components. Green indicates positive values, yellow indicates negative values. A detailed description of the components can be found in Table S7. Created with IBM SPSS Statistics and MS Office.

Table S1. MS/MS parameters for selected compounds.

| Analytes | Tr | Q1 (*m/z*) | Q3 (*m/z*) | DP (V) | CE (V) | CXP (V) |
|---|---|---|---|---|---|---|
| **ESI(+)** | | | | | | |
| **IMP** | 0.98 | 300.0 | 98.1/142.1 | 66 | 41/39 | 16/8 |
| **MEM** | 2.82 | 384.2 | 68.1/141.1 | 61 | 67/25 | 10/10 |
| **ESI(-)** | | | | | | |
| **VBR** | 3.46 | 296.0 | 234.1/278.2 | -65 | -28/-20 | -7/-7 |
| **CIL** | 3.50 | 357.0 | 226.3/313.2 | -60 | -26/-22 | -5/-13 |

tr - retention time; Q1 - precursor ion; Q3 - fragment ion; DP - declustering potential; CE - collision energy; CXP - cell exit potential; IMI – imipenem; MEM – meropenem; VBR – vaborbactam; CIL – cilstatin

Table S2: Basic information on chromatographic analysis of selected anions.

| Anion | LOQ [mg/L] | Retention time [min] |
|---|---|---|
| $F^-$ | 0.10 | 4.0 ± 0.15 |
| $Cl^-$ | 0.15 | 6.2 ± 0.20 |
| $NO_2^-$ | 0.50 | 7.3 ± 0.25 |
| $Br^-$ | 0.50 | 9.0 ± 0.20 |
| $NO_3^-$ | 0.50 | 10.3 ± 0.25 |
| $PO_4^{3-}$ | 0.75 | 14.3 ± 0.20 |
| $SO_4^{2-}$ | 0.75 | 15.9 ± 0.25 |

Table S3: Primers used in multiplex PCR reactions for identification of *Acinetobacter baumannii* and *Pseudomonas aeruginosa*.

| Primer | 3′→ 5′primer sequence | Product size [bp] | Annealing temperature [°C] | References |
|---|---|---|---|---|
| ***Ab*-ITS** | F: CATTATCACGGTAATTAGTG<br>R: AGAGCACTGTGCACTTAAG | 208 | 55 | (1,2) |
| ***opr*L** | F: ATGGAAATGCTGAAATTCGGC<br>R: CTTCTTCAGCTCGACGCGACG | 504 | 57 | (3,4) |

Table S4: Primers used in PCR, multiplex PCR and ERIC PCR reaction for identification of antibiotic resistance genes in obtained *Acinetobacter baumannii* and *Pseudomonas aeruginosa*

| Primer | 3'→ 5'primer sequence | Product size [bp] | Annealing temperature [°C] | References |
|---|---|---|---|---|
| ***bla***$_{\text{OXA 23-like}}$ | F: GATCGGATTGGAGAACCAGA<br>R: ATTTCTGACCGCATTTCCAT | 501 | 53 | |
| ***bla***$_{\text{OXA 24-like}}$ | F: GGTTAGTTGGCCCCCTTAAA<br>R: AGTTGAGCGAAAAGGGGATT | 246 | 53 | |
| ***bla***$_{\text{OXA 51-like}}$ | F: TAATGCTTTGATCGGCCTTG<br>R: TGGATTGCACTTCATCTTGG | 353 | 53 | (5) |
| ***bla***$_{\text{OXA 58-like}}$ | F: AAGTATTGGGGCTTGTGCTG<br>R: CCCCTCTGCGCTCTACATAC | 599 | 53 | |
| ***bla***$_{\text{VIM}}$ | F: TTTGGTCGCATATCGCAACG<br>R: CCATTCAGCCAGATCGGCAT | 500 | 53 | |
| ***bla***$_{\text{BIC}}$ | F: TATGCAGCTCCTTTAAGGGC<br>R: TCATTGGCGGTGCCGTACAC | 537 | 52 | |
| ***bla***$_{\text{SPM}}$ | F: AAAATCTGGGTACGCAAACG<br>R: ACATTATCCGCTGGAACAGG | 271 | 52 | (6) |
| ***bla***$_{\text{AIM}}$ | F: CTGAAGGTGTACGGAAACAC<br>R: GTTCGGCCACCTCGAATTG | 322 | 52 | |
| ***bla***$_{\text{GIM}}$ | F: TCGACACACCTTGGTCTGAA<br>R: AACTTCCAACTTTGCCATGC | 477 | 52 | |
| ***bla***$_{\text{CTX}}$ | F: GACAAAGAGAGTGCAACGGATG<br>R: TCAGTGCGATCCAGACGAAA | 501 | 61 | |
| ***bla***$_{\text{OXA}}$ | F: ATTATCTACAGCAGCGCCAGTG<br>R: TGCATCCACGTCTTTGGTG | 296 | 61 | (7) |
| ***bla***$_{\text{TEM}}$ | F: AGTGCTGCCATAACCATGAGTG<br>R: CTGACTCCCC GTCGTGTAGATA | 431 | 61 | |
| ***bla***$_{\text{SHV}}$ | F: GATGAACGCTTTCCCATGATG<br>R: CGCTGTTATCGCTCATGGTAA | 214 | 61 | |
| ***tet*A** | F: GCTACATCCTGCTTGCCTTC<br>R: GCATAGATCGCCGTGAAGAG | 211 | 54 | (8) |
| ***tet*M** | F: ACAGAAAGCTTATTATATAAC<br>R: TGGCGTGTCTATGATGTTCAC | 171 | 55 | |
| ***qep*A** | F: GCCGGTGATGCTGCTGA<br>R: CAGTAACAGCGCACCAA | 204 | 63 | (9) |
| ***erm*F** | F: TAGATATTGGGGCAGGCAAG<br>R: GGAAATTGCGGAACTGCAAA | 126 | 58 | (10) |
| ***sul*1** | F: CGCACCGGAAACATCGCTGCAC<br>R: TGAAGTTCCGCCGCAAGGCTCG | 163 | 56 | (11) |
| ***sul*2** | F: TCCGGTGGAGGCCGGTATCTGG<br>R: CGGGAATGCCATCTGCCTTGAG | 191 | 57.5 | |
| ***bla***$_{\text{VIM}}$ | F: GAGTTGCTTTTGATTGATACAG<br>R: TCGATGAGAGTCCTTCTAGA | 247 | 50 | |
| ***bla***$_{\text{IMP}}$ | F: TTGACACTCCATTTACTGCTA<br>R: TCATTTGTTAATTCAGATGCATA | 172 | 50 | (12) |
| ***bla***$_{\text{NDM}}$ | F: AACACAGCCTGACTTTCG<br>R: TGATATTGTCACTGGTGTGG | 111 | 50 | |
| **ERIC 1**<br>**ERIC 2** | F: ATGTAAGCTCCTGGGGATTCAC<br>R: AAGTAAGTGACTGGGGTGAGCG | 50-420 | 52/65 | (13,14) |

Table S5: Ranges of antibiotics concentration in samples of hospital wastewater collected in both seasons across Poland

| **Season** | **Voivodeship** | **Concentrations of tested antibiotics** | | |
|---|---|---|---|---|
| | | **Meropenem** MIN-MAX [ng/L] (MEAN ± SD*) | **Cilastatin/Imipenem** MIN-MAX [ng/L] (MEAN ± SD) | **Vaborbactam** MIN-MAX [ng/L] (MEAN ± SD) |
| **Winter** | Lower Silesia | 0.0 – 138.59 (71.13 ± **95.41**) | 6.00 – 27.91 (10.58 ± **11.56**) | 0.00 – 26.60 (26.60 ± **2.11**) |
| | Kuyavian–Pomerania | - | 2.95 – 6.55 (3.12 ± **2.03**) | 0.00 – 32.68 (22.91 ± **13.81**) |
| | Lublin | 0.00 – 84.18 (84.18 ± **8.11**) | 6.85 – 16.79 (11.96 ± **5.59**) | 0.00 – 66.48 (54.30 ± **17.23**) |
| | Lubusz | 0.00 – 23.85 (23.85 ± **2.75**) | 11.82 – 26.74 (16.96 ± **7.58**) | 0.00 – 85.13 (85.13 ± **3.93**) |
| | Łódź | 0.00 – 6.22 (6.22 ± **1.06**) | 3.45 – 13751.116 (10.90 ± **7935.09**) | 0.00 – 62.75 (51.58 ± **15.80**) |
| | Małopolska | 0.00 – 176.23 (77.28 ± **82.62**) | 0.00 – 9.35 (6.80 ± **2.66**) | 0.00 – 93.00 (22.88 ± **41.51**) |
| | Mazovia | 0.00 – 23.08 (13.91 ± **12.97**) | 0.00 – 17315.79 (110.37 ± **9958.82**) | 0.00 – 51.42 (31.13 ± **28.60**) |
| | Opole | 0.00 – 49.71 (49.71 ± **4.88** | 0.00 – 4343.45 (11.98 ± **2501.25**) | 0.00 – 51.52 (40.83 ± **14.98**) |
| | Podkarpacie | 0.00 – 102.67 (62.50 ± **56.80**) | 6.83 – 4718.82 (9.92 ± **2354.96**) | 0.00 – 39.76 (38.41 ± **1.92**) |
| | Podlasie | 0.00 – 5.57 (5.57 ± **0.49**) | 0.00 – 22.36 (19.15 ± **4.54**) | 0.00 – 39.76 (49.77 ± **3.90**) |
| | Pomerania | 0.00 – 15.20 (15.20 ± **2.09**) | 6.91 – 29.95 (11.41 ± **9.34**) | 0.00 – 48.97 (37.36 ± **16.42**) |
| | Silesia | 0.00 – 163.23 (163.23 ± **14.64**) | 0.00 – 25.91 (5.11 ± **12.87**) | 0.00 – 21.08 (21.08 ± **0.62**) |
| | Świętokrzyskie | 3.57 – 45.01 (16.65 ± **19.88**) | 5.89 – 17.91 (6.48 ± **5.82**) | 0.00 – 38.45 (22.54 ± **22.50**) |
| | Warmia-Masuria | 0.00-13.86 (13.86 ± **0.42**) | 0.00 – 2375.24 (2375.24 ± **277.80**) | - |
| | Wielkopolska | 0.00 – 2.69 (2.69 ± **0.25**) | 5.51 – 16.63 (8.74 ± **4.85**) | 0.00 – 59.29 (59.29 ± **2.41**) |
| | West Pomerania | 0.00 – 7.76 (7.76 ± **0.80**) | 4.07 – 9.69 (7.53 ± **2.40**) | 0.00 – 35.51 (30.94 ± **6.47**) |
| **Summer** | Lower Silesia | 1.49 – 377.38 (199.23 ± **188.03**) | 8.45 – 10.78 (10.70 ± **1.32**) | - |
| | Kuyavian–Pomerania | 20.65 – 79.46 (22.34 ± **33.48**) | 5.62 – 13.13 (8.90 ± **3.77**) | - |
| | Lublin | 10.51 – 351.52 (94.79 ± **145.05**) | 0.00 – 7.87 (3.73 ± **2.49**) | 11.29 – 48.93 (12.56 ± **18.17**) |
| | Lubusz | 6.40 – 128.44 (80.59 ± **526.70**) | 15.27 – 51.98 (12.33 ± **4.33**) | 0.00 – 14.40 (8.30 ± **1.10**) |
| | Łódź | 6.40 – 128.44 (9.16 ± **56.94**) | 15.27 – 51.98 (15.27 ± **18.18**) | 0.00 – 14.40 (9.05 ± **3.78**) |
| | Małopolska | 0.00 – 40.12 (6.23 ± **19.90**) | 0.00 – 11.17 (4.75 ± **3.95**) | 0.00 – 9.13 (7.89 ± **1.75**) |
| | Mazovia | 0.00 – 30.65 (30.65 ± **4.80**) | 2.85 – 294.37 (9.34 ± **143.06**) | 0.00 – 16.33 (16.33 ± **1.70**) |
| | Opole | 0.00 – 15.62 (10.98 ± **6.57**) | 1.61 – 16.08 (9.49 ± **6.11**) | - |
| | Podkarpacie | 14.68 – 102.56 (16.39 ± **43.00**) | 0.00 – 20.68 (11.63 ± **5.75**) | 0.00 – 13.25 (13.25 ± **1.38**) |
| | Podlasie | 0.00 – 81.64 (55.84 ± **18.24**) | 3.12 – 12.88 (9.69 ± **4.17**) | - |
| | Pomerania | 0.00 – 25.97 (15.13 ± **15.33**) | 4.07 – 10.65 (6.82 ± **2.63**) | 0.00 – 10.62 (10.62 ± **1.66**) |
| | Silesia | 0.00 – 333.41 (201.50 ± **167.11**) | 1.20 – 5.97 (2.50 ± **2.02**) | 0.00 – 26.28 (15.98 ± **7.28**) |
| | Świętokrzyskie | 0.00 – 26.74 (25.94 ± **1.14**) | 5.49 – 14.63 (11.80 ± **4.05**) | 0.00 – 9.54 (4.66 ± **3.45**) |
| | Warmia-Masuria | 0.00 – 171.08 (27.33 ± **101.65**) | 6.15 – 2180.12 (9.03 ± **1045.89**) | 0.00 32.49 (5.13 ± **19.35**) |
| | Wielkopolska | 9.35 – 428.15 (29.71 ± **201.26**) | 2.72 – 17.78 (4.72 ± **6.83**) | 0.00 – 52.63 (7.54 ± **31.89**) |
| | West Pomerania | 0.00 – 47.95 (6.03 ± **24.49**) | 0.00 – 14.27 (7.11 ± **5.06**) | 0.00 – 104.96 (35.08 ± **46.41**) |

*SD: standard deviation

## Table S6: Chemical parameters of samples of hospital wastewater collected in both seasons across Poland.

| Season | Voivodeship | Median concentrations of tested parameters | | | | | | | | | | | |
|---|---|---|---|---|---|---|---|---|---|---|---|---|---|
| | | **TOC [mg/L] ± SD*** | **TN [mg/L] ± SD** | **$N-NH_4^+$ [mg/L] ± SD** | **N-org [mg/L] ± SD** | **pH ± SD** | **$F^-$ [mg/L] ± SD** | **$Cl^-$ [mg/L] ± SD** | **$N-NO^{2-}$ [mg/L] ± SD** | **$Br^-$ [mg/L] ± SD** | **$N-NO_3^-$ [mg/L] ± SD** | **$PO_4^{3-}$ [mg/L] ± SD** | **$SO_4^{2-}$ [mg/L] ± SD** |
| **Winter** | Lower Silesia | 76.10 ± **3.76** | 17.74 ± **7.28** | 3.21 ± **15.27** | 13.33 ± **8.06** | 7.00 ± **0.50** | 0.56 ± **0.05** | 116.41 ± **31.08** | 0.59 ± **0.54** | 0.00 ± **1.13** | 1.53 ± **1.18** | 7.89 ± **5.97** | 79.47 ± **16.18** |
| | Kuyavian–Pomerania | 68.00 ± **21.08** | 18.55 ± **2.88** | 14.00 ± **1.84** | 4.25 ± **3.55** | 7.00 ± **0.50** | 0.41 ± **0.16** | 436.69 ± **242.15** | 0.00 ± **0.65** | 0.62 ± **0.53** | 1.12 ± **0.51** | 12.70 ± **7.42** | 75.31 ± **140.41** |
| | Lublin | 209 ± **80.93** | 21.82 ± **6.63** | 17.93 ± **4.73** | 3.56 ± **2.63** | 7.25 ± **0.48** | 0.77 ± **0.13** | 270.48 ± **228.68** | 0.00 ± **0.61** | 0.24 ± **0.28** | 1.87 ± **2.51** | 11.11 ± **2.72** | 39.12 ± **12.86** |
| | Lubusz | 288.80 ± **184.15** | 19.88 ± **12.98** | 16.40 ± **13.22** | 1.17 ± **1.77** | 6.50 ± **0.00** | 0.56 ± **4.59** | 169.20 ± **113.25** | - | - | 0.00 ± **0.86** | 16.24 ± **10.85** | 66.88 ± **70.61** |
| | Łódź | 161.43 ± **47.16** | 18.20 ± **6.95** | 13.03 ± **6.11** | 7.59 ± **5.19** | 6.75 ± **0.65** | 1.18 ± **8.68** | 205.37 ± **359.96** | 0.00 ± **0.27** | 0.22 ± **0.30** | 1.33 ± **0.70** | 15.29 ± **10.51** | 38.85 ± **49.06** |
| | Małopolska | 111.25 ± **38.38** | 26.34 ± **11.41** | 7.64 ± **6.89** | 6.79 ± **14.36** | 8.00 ± **0.96** | 0.69 ± **0.22** | 216.18 ± **107.80** | 0.55 ± **1.03** | 0.00 ± **0.24** | 0.00 ± **1.57** | 5.16 ± **5.73** | 82.52 ± **28.25** |
| | Mazovia | 108.93 ± **74.07** | 23.16 ± **5.92** | 2.81 ± **7.08** | 15.25 ± **5.48** | 7.50 ± **0.75** | 0.53 ± **0.23** | 91.50 ± **120.99** | - | 0.72 ± **1.02** | 0.00 ± **0.78** | 8.16 ± **7.37** | 74.66 ± **18.42** |
| | Opole | 132.70 ± **71.07** | 20.26 ± **7.96** | 7.02 ± **10.02** | 5.53 ± **10.90** | 7.00 ± **0.67** | 0.39 ± **0.12** | 117.63 ± **61.13** | 0.54 ± **1.26** | 0.00 ± **0.51** | 0.41 ± **2.03** | 4.41 ± **6.24** | 111.50 ± **54.63** |
| | Podkarpacie | 168.63 ± **129.82** | 21.96 ± **22.18** | 20.08 ± **22.73** | 0.00 ± **1.88** | 7.25 ± **0.95** | 0.50 ± **0.47** | 185.39 ± **184.68** | 0.00 ± **0.55** | 0.00 ± **0.27** | 0.75 ± **0.84** | 9.66 ± **6.13** | 81.69 ± **48.61** |
| | Podlasie | 64.72 ± **79.97** | 9.08 ± **3.09** | 8.85 ± **3.34** | 0.23 ± **0.37** | 6.75 ± **0.61** | 0.51 ± **2.30** | 72.44 ± **132.50** | 0.00 ± **0.33** | - | 0.49 ± **0.68** | 2.31 ± **6.98** | 51.16 ± **23.54** |
| | Pomerania | 71.35 ± **52.94** | 21.22 ± **18.23** | 2.99 ± **8.62** | 9.68 ± **19.82** | 6.50 ± **0.61** | 0.52 ± **0.29** | 5.02 ± **238.55** | 1.14 ± **0.52** | - | 1.62 ± **2.52** | 15.53 ± **9.12** | 58.62 ± **68.77** |
| | Silesia | 216.48 ± **155.92** | 14.65 ± **8.28** | 7.48 ± **6.58** | 2.63 ± **6.70** | 7.50 ± **0.84** | 0.40 ± **0.21** | 96.77 ± **74.33** | 0.00 ± **1.26** | 0.00 ± **0.23** | 0.43 ± **2.11** | 2.42 ± **3.42** | 63.68 ± **154.29** |
| | Świętokrzyskie | 161.15 ± **51.55** | 20.39 ± **6.58** | 10.13 ± **11.66** | 6.12 ± **7.01** | 6.50 ± **0.25** | 0.52 ± **0.68** | 164.84 ± **103.43** | 0.97 ± **0.70** | 0.00 ± **0.23** | 1.26 ± **2.91** | 13.99 ± **7.85** | 44.74 ± **8.83** |
| | Warmia-Masuria | 146.28 ± **121.46** | 13.99 ± **10.32** | 13.99 ± **10.43** | 0.00 ± **0.13** | 7.25 ± **0.61** | 0.50 ± **0.18** | 151.29 ± **118.57** | 0.57 ± **1.69** | - | 1.54 ± **8.54** | 9.19 ± **20.64** | 51.32 ± **11.98** |
| | Wielkopolska | 216.33 ± **134.68** | 19.74 ± **9.54** | 17.95 ± **6.94** | 1.79 ± **3.53** | 7.00 ± **0.58** | 0.51 ± **7.41** | 396.99 ± **340.19** | - | 0.51 ± **0.26** | 1.31 ± **0.79** | 14.11 ± **7.01** | 79.80 ± **46.38** |
| | West Pomerania | 137.38 ± **152.62** | 15.93 ± **8.75** | 13.50 ± **8.66** | 0.00 ± **2.44** | 6.75 ± **0.65** | 0.48 ± **0.18** | 61.34 ± **38.28** | 0.00 ± **0.27** | 0.00 ± **0.98** | - | 8.10 ± **5.95** | 53.21 ± **70.33** |
| **Summer** | Lower Silesia | 152.20 ± **87.04** | 33.87 ± **13.85** | 30.55 ± **13.96** | 0.00 ± **1.92** | 7.00 ± **0.17** | 0.48 ± **0.11** | 75.65 ± **25.17** | - | 0.00 ± **0.32** | - | 9.16 ± **8.22** | 67.55 ± **40.36** |
| | Kuyavian–Pomerania | 169.15 ± **31.65** | 14.18 ± **9.17** | 14.18 ± **9.17** | - | 7.00 ± **0.56** | 0.43 ± **0.28** | 593.21 ± **273.47** | 0.00 ± **0.68** | 0.50 ± **0.68** | 0.51 ± **0.43** | 5.57 ± **5.86** | 101.54 ± **86.35** |
| | Lublin | 208.30 ± **130.78** | 25.05 ± **8.21** | 24.63 ± **7.94** | 0.00 ± **0.60** | 6.65 ± **0.24** | 0.49 ± **0.12** | 185.94 ± **160.45** | - | - | - | 16.59 ± **5.53** | 29.41 ± **17.54** |
| | Lubusz | 163.50 ± **104.77** | 22.39 ± **1.93** | 22.39 ± **1.93** | - | 6.50 ± **0.12** | 0.47 ± **0.09** | 395.71 ± **87.31** | 0.00 ± **0.85** | 0.00 ± **0.61** | 1.76 ± **1.17** | 18.94 ± **1.62** | 35.61 ± **84.65** |
| | Łódź | 179.48 ± **225.18** | 19.65 ± **8.54** | 19.65 ± **8.54** | - | 6.60 ± **0.33** | 0.46 ± **0.07** | 212.86 ± **463.35** | 0.00 ± **0.62** | 0.23 ± **1.12** | 0.30 ± **0.90** | 5.34 ± **2.77** | 27.74 ± **9.54** |
| | Małopolska | 92.25 ± **138.02** | 11.77 ± **14.52** | 11.77 ± **8.16** | 0.00 ± **7.87** | 6.50 ± **0.37** | 0.32 ± **0.07** | 492.09 ± **338.96** | 0.00 ± **1.24** | 0.48 ± **0.28** | - | 8.22 ± **6.71** | 67.78 ± **67.75** |
| | Mazovia | 235.90 ± **228.87** | 23.98 ± **14.96** | 23.98 ± **17.98** | 0.00 ± **4.83** | 6.95 ± **1.46** | 0.72 ± **38.30** | 231.19 ± **153.65** | 0.29 ± **0.72** | 0.00 ± **1.09** | - | 4.02 ± **5.07** | 19.28 ± **24.42** |
| | Opole | 175.90 ± **53.77** | 29.27 ± **12.34** | 23.19 ± **14.48** | 0.00 ± **6.36** | 7.20 ± **0.57** | 0.41 ± **0.18** | 137.57 ± **382.75** | 0.00 ± **0.24** | 0.00 ± **0.21** | - | 12.63 ± **9.29** | 24.76 ± **24.12** |
| | Podkarpacie | 119.98 ± **44.69** | 13.99 ± **8.74** | 13.60 ± **8.82** | 0.00 ± **0.39** | 6.65 ± **0.38** | 0.50 ± **0.30** | 211.89 ± **200.24** | 0.00 ± **5.29** | - | - | 3.69 ± **6.19** | 19.13 ± **64.06** |
| | Podlasie | 177.83 ± **197.06** | 15.93 ± **7.84** | 15.93 ± **9.45** | 0.00 ± **1.80** | 6.90 ± **0.53** | 0.48 ± **0.18** | 125.85 ± **282.13** | 0.54 ± **0.88** | 0.00 ± **0.25** | 1.11 ± **1.90** | 3.02 ± **2.56** | 35.91 ± **15.91** |
| | Pomerania | 178.25 ± **99.56** | 21.57 ± **32.91** | 21.57 ± **32.95** | 0.00 ± **0.16** | 6.50 ± **0.96** | 0.56 ± **0.49** | 216.72 ± **173.98** | 0.00 ± **0.24** | - | - | 10.57 ± **7.70** | 53.28 ± **34.55** |
| | Silesia | 78.93 ± **111.82** | 18.72 ± **6.99** | 18.72 ± **6.99** | - | 6.50 ± **0.50** | 0.36 ± **18.10** | 96.83 ± **35.29** | 0.00 ± **0.62** | 0.00 ± **0.52** | - | 9.41 ± **4.35** | 60.97 ± **48.48** |
| | Świętokrzyskie | 214.08 ± **117.43** | 13.37 ± **8.37** | 13.37 ± **9.16** | 3.42 ± **5.36** | 6.50 ± **0.49** | 0.38 ± **26.30** | 386.61 ± **260.84** | - | 0.00 ± **0.26** | 0.00 ± **0.81** | 0.00 ± **58.61** | 0.00 ± **12.17** |
| | Warmia-Masuria | 168.95 ± **119.48** | 22.87 ± **2.14** | 22.87 ± **2.14** | - | 7.25 ± **0.39** | 0.67 ± **0.23** | 157.64 ± **113.27** | 0.54 ± **0.44** | 0.24 ± **0.28** | - | 7.95 ± **6.98** | 55.11 ± **15.13** |
| | Wielkopolska | 169.78 ± **368.43** | 26.75 ± **15.03** | 26.75 ± **15.03** | - | 6.85 ± **0.92** | 0.41 ± **57.60** | 409.38 ± **79.15** | 0.64 ± **0.84** | 0.23 ± **0.48** | 1.00 ± **1.64** | 15.30 ± **15.94** | 69.43 ± **59.74** |
| | West Pomerania | 192.35 ± **77.36** | 18.15 ± **4.81** | 18.15 ± **4.81** | - | 6.50 ± **0.25** | 0.40 ± **0.15** | 123.20 ± **66.13** | 0.00 ± **0.28** | 0.22 ± **0.26** | - | 11.61 ± **8.49** | 94.29 ± **74.31** |

*SD: standard deviation; TOC: Total organic carbon; TN: Total nitrogen; N-org: organic nitrogen;

Table S7: Mean number of obtained cream-colored CR colonies (CCRC) vs all CR colonies (ACRC) in HWW samples across voivodeships in Poland.

| Season | Voivodeship | Mean of CCRC in 1 mL [cfu/mL] | Mean of ACRC in 1 mL [cfu/mL] | CCRC among ACRC [%] |
|---|---|---|---|---|
| **Winter** | Lower Silesia | 2.E+05 | 8.E+05 | 21 |
| | Kuyavian–Pomerania | 4.E+04 | 1.E+05 | 25 |
| | Lublin | 5.E+04 | 2.E+05 | 23 |
| | Lubusz | 4.E+04 | 2.E+05 | 23 |
| | Łódź | 9.E+04 | 4.E+05 | 24 |
| | Małopolska | 4.E+04 | 3.E+05 | 13 |
| | Mazovia | 3.E+03 | 5.E+03 | 61 |
| | Opole | 5.E+04 | 4.E+05 | 12 |
| | Podkarpacie | 7.E+04 | 3.E+05 | 27 |
| | Podlasie | 6.E+04 | 1.E+05 | 40 |
| | Pomerania | 8.E+03 | 2.E+04 | 41 |
| | Silesia | 3.E+04 | 1.E+05 | 22 |
| | Świętokrzyskie | 4.E+04 | 2.E+05 | 16 |
| | Warmia-Masuria | 3.E+03 | 1.E+04 | 21 |
| | Wielkopolska | 6.E+03 | 4.E+04 | 14 |
| | West Pomerania | 2.E+04 | 5.E+04 | 37 |
| **Summer** | Lower Silesia | 2.E+05 | 2.E+06 | 9 |
| | Kuyavian–Pomerania | 5.E+04 | 5.E+05 | 10 |
| | Lublin | 1.E+05 | 9.E+05 | 10 |
| | Lubusz | 2.E+04 | 6.E+05 | 3 |
| | Łódź | 9.E+04 | 7.E+05 | 14 |
| | Małopolska | 4.E+05 | 1.E+06 | 25 |
| | Mazovia | 1.E+05 | 8.E+05 | 13 |
| | Opole | 9.E+05 | 3.E+06 | 35 |
| | Podkarpacie | 1.E+05 | 7.E+05 | 20 |
| | Podlasie | 5.E+04 | 3.E+05 | 17 |
| | Pomerania | 1.E+04 | 2.E+05 | 5 |
| | Silesia | 1.E+05 | 9.E+05 | 12 |
| | Świętokrzyskie | 7.E+04 | 3.E+05 | 28 |
| | Warmia-Masuria | 1.E+04 | 2.E+05 | 6 |
| | Wielkopolska | 2.E+05 | 7.E+05 | 25 |
| | West Pomerania | 3.E+04 | 4.E+05 | 6 |

Table S8. PCA main components description

| Component | Main variables (charges ≥ 0.6) |
|---|---|
| PC1 | N-$NH_4^+$ (0,906), TN (0,833), $F^-$ (0,785), pH (–0,759), TOC (0,701) |
| PC2 | Volume of wastewater (0,691), Number of beds (0,653), CCRC* (0,541), ACCAH*** (0,581) |
| PC3 | ACRC** (0,728), $Cl^-$ (0,734), Varbobactam (0,629) |
| PC4 | $Po_4^{3-}$ (0,723), Varbobactam (0,629), $SO_4^{2-}$ (0,409), TOC (0,243) |
| PC5 | $Cl^-$ (–0,495), Meropenem (0,723), Varbobactam (–0,390) |
| PC6 | $NO_3^-$ (0,860), $NO_2^-$ (0,580), Volume (–0,275) |
| PC7 | N-org (0,523), TOC (0,329), Cilstatin (0,273) |

*CCRC – Cream Carbapenem Resistant Colonies
**ACRC – All Carbapenem Resistant Colonies
***ACCAH - annual consumption of carbapenem antibiotics in hospital

## References

1. Hubeny J, Korzeniewska E, Buta-Hubeny M, Zieliński W, Rolbiecki D, Harnisz M. Characterization of carbapenem resistance in environmental samples and Acinetobacter spp. isolates from wastewater and river water in Poland. Sci Total Environ. 2022;822.

2. Chen TL, Siu LK, Wu RCC, Shaio MF, Huang LY, Fung CP, et al. Comparison of one-tube multiplex PCR, automated ribotyping and intergenic spacer (ITS) sequencing for rapid identification of Acinetobacter baumannii. Clin Microbiol Infect. 2007;13(8):801–6.

3. De Vos D, Lim A, Pirnay JP, Struelens M, Vandenvelde C, Duinslaeger L, et al. Direct detection and identification of Pseudomonas aeruginosa in clinical samples such as skin biopsy specimens and expectorations by multiplex PCR based on two outer membrane lipoprotein genes, oprI and oprL. J Clin Microbiol. 1997;35(6):1295–9.

4. Al-Ahmadi J, Roodsari Z. Fast and Specific Detection of Pseudomonas Aeruginosa From Other Pseudomonas Species By Pcr. Ann Burns Fire Disasters [Internet]. 2016;XXIX(December):5–8. Available from: http://www.ncbi.nlm.nih.gov/pubmed/28289359%0Ahttp://www.pubmedcentral.nih.gov/articlerender.fcgi?artid=PMC5347312

5. Amudhan SM, Sekar U, Arunagiri K, Sekar B. OXA betaβ-lactamase-mediated carbapenem resistance in Acinetobacter baumannii. Indian J Med Microbiol [Internet]. 2011;29(3):269–74. Available from: https://doi.org/10.4103/0255-0857.83911

6. Poirel L, Walsh TR, Cuvillier V, Nordmann P. Multiplex PCR for detection of acquired carbapenemase genes. Diagn Microbiol Infect Dis [Internet]. 2011;70(1):119–23. Available from: http://dx.doi.org/10.1016/j.diagmicrobio.2010.12.002

7. Kim J, Jeon S, Rhie H, Lee B, Park M, Lee H, et al. Rapid Detection of Extended Spectrum β-Lactamase (ESBL) for Enterobacteriaceae by use of a Multiplex PCR-based Method. Infect Chemother. 2009;41(3):181.

8. Ng LK, Martin I, Alfa M, Mulvey M. Multiplex PCR for the detection of tetracycline resistant genes. Mol Cell Probes. 2001;15(4):209–15.

9. Goldstein C, Lee MD, Sanchez S, Hudson C, Phillips B, Register B, et al. Incidence of class 1 and 2 integrases in clinical and commensal bacteria from livestock, companion animals, and exotics. Antimicrob Agents Chemother. 2001;45(3):723–6.

10. Eitel Z, Sóki J, Urbán E, Nagy E. The prevalence of antibiotic resistance genes in Bacteroides fragilis group strains isolated in different European countries. Anaerobe. 2013;21:43–9.

11. Pei R, Kim SC, Carlson KH, Pruden A. Effect of River Landscape on the sediment concentrations of antibiotics and corresponding antibiotic resistance genes (ARG). Water Res. 2006;40(12):2427–35.

12. Goudarzi H, Mirsamadi ES, Ghalavand Z, Vala MH, Mirjalali H, Hashemi A, et al. Molecular detection of metallo-beta-lactamase genes in clinical isolates of Acinetobacter baumannii. J Pure Appl Microbiol. 2015;9(Special Edition 2):145–51.

13. Versalovic J, Koeuth T, Lupski R. Distribution of repetitive DNA sequences in eubacteria and application to finerpriting of bacterial enomes. Nucleic Acids Res. 1991;19(24):6823–31.

14. Osińska A, Korzeniewska E, Harnisz M, Niestępski S. The prevalence and characterization of antibiotic-resistant and virulent Escherichia coli strains in the municipal wastewater system and their environmental fate. Sci Total Environ. 2017;577:367–75.